\documentclass[iop]{emulateapj}  
 
\usepackage[english]{babel}
\usepackage{amsmath}
\usepackage{graphicx}
\usepackage[dvipsnames]{xcolor}
\usepackage{parskip}
\usepackage{wrapfig}
\usepackage{url}
\usepackage{float}
\usepackage{color}
\usepackage{amssymb}
\usepackage{multirow}

\newcommand{\mum}{\ifmmode{\rm \mu m}\else{$\mu$m }\fi}

\newcommand{\chisq}{\ifmmode{\chi^{2} }\else{$\chi^2$}\fi}
\newcommand{\rchisq}{\ifmmode{\chi^{2} }\else{$\chi^2_\nu$}\fi}

\newcommand{\Hii}{H {\sc ii} }

\begin{document}

\shorttitle{Mid-IR stellar populations with JWST/MIRI}  
\shortauthors{Jones et al.}

\title{Probing the dusty stellar populations of the Local Volume Galaxies with JWST/MIRI}

\author{Olivia~C.~Jones\altaffilmark{1},
        Margaret~Meixner\altaffilmark{1,2},
        Kay Justtanont\altaffilmark{3} 
\& 	Alistair Glasse\altaffilmark{4} 
        }
\altaffiltext{1}{Space Telescope Science Institute, 3700 San Martin Drive, Baltimore, MD, 21218, USA}
\altaffiltext{2}{The Johns Hopkins University, Department of Physics and Astronomy, 366 Bloomberg Center, 3400 N. Charles Street, Baltimore, MD 21218, USA}
\altaffiltext{3}{Department of Earth \& Space Sciences, Chalmers University of Technology, Onsala Space Observatory, SE-439 92 Onsala, Sweden} 
\altaffiltext{4}{UK Astronomy Technology Centre, Royal Observatory, Edinburgh, Blackford Hill, Edinburgh EH9 3HJ, UK}


\begin{abstract}
  The Mid-Infrared Instrument (MIRI) for the {\em James Webb Space Telescope} (JWST) will revolutionize our understanding of infrared stellar populations in the Local Volume. Using the rich {\em Spitzer}-IRS spectroscopic data-set and spectral classifications from the Surveying the Agents of Galaxy Evolution (SAGE)-Spectroscopic survey of over a thousand objects in the Magellanic Clouds,  the Grid of Red supergiant and Asymptotic giant branch star ModelS ({\sc grams}), and the grid of YSO models by \citet{Robitaille2006}, we calculate the expected flux-densities and colors in the MIRI broadband filters for prominent infrared stellar populations. We use these fluxes to explore the {\em JWST}/MIRI colours and magnitudes for composite stellar population studies of Local Volume galaxies.
  MIRI colour classification schemes are presented; these diagrams provide a powerful means of identifying young stellar objects, evolved stars and extragalactic background galaxies in Local Volume galaxies with a high degree of confidence. 
Finally, we examine which filter combinations are best for selecting populations of sources based on their JWST colours.
\end{abstract}

\keywords{stars: late-type  -- infrared: stars  -- Magellanic Clouds }

\section{Introduction} \label{sec:intro}

The {\em James Webb Space Telescope} (JWST) will revolutionize our knowledge of the stellar populations of galaxies out to and beyond the Virgo cluster.  The large (6.5~m) aperture of {\em JWST} combined with its subarcsecond spatial resolution and broad wavelength coverage (0.6--28.5 $\mu$m) will provide unprecedented opportunities to study the properties of resolved dusty stellar populations at moderate and large distances ($\sim$4 Mpc). Notably, {\em JWST} will enable the accurate measurement of the star formation histories of nearby galaxies, determine the age of the oldest stellar populations and characterize metal enrichment in a wide variety of environments. 

Infrared observations of stellar populations are critical for studies of chemical enrichment, as both the early and late stages of a star's life are enshrouded in dust and molecular gas. However, the ability to correctly interpret infrared observations (especially for galaxies at high-redshift) relies on accurate constraints of the stellar populations.
Currently, resolved mid-infrared stellar population studies of galaxies with multiple populations are limited to nearby galaxies such as the Magellanic Clouds.  With {\em JWST} this will be possible for well over a hundred galaxies within the Local Volume. 

The dusty stellar populations of the Large and Small Magellanic Clouds (D = 50 and 60 kpc; \citealt{Ngeow2008, Szewczyk2009, Feast2013}), with metallicities of 0.5 and 0.2 $Z_{\odot}$ \citep{Russell1992}, have been well characterized in the IR from 3.6--500 $\mu$m using the {\em Spitzer} and {\em Herschel} Space Telescopes \citep{Meixner2006, Meixner2013, Gordon2011}. Over 10.5 million point sources were photometrically detected, and key populations of sources throughout the Magellanic Clouds identified in the colour magnitude diagrams (CMDs). Furthermore, over 1250 of these sources were observed with {\em Spitzer's} InfraRed Spectrograph (IRS), revealing the mineralogy and evolutionary status of these sources \citep[][Jones et al. in press.]{Woods2011, Ruffle2015}. This large spectroscopic data-set and corresponding point source classification is an ideal empirical library of template spectrum, which we can use to predict the flux densities and magnitudes that we will detect with the {\em JWST} Mid-Infrared Instrument (MIRI; \citealt{Rieke2015}). 
  
The Mid-Infrared Instrument (MIRI; \citealt{Rieke2015}) on board {\em JWST} will provide broad-band imaging, coronography and spectroscopy over the 5--28.5 $\mu$m wavelength range \citep{Rieke2015, Wright2015, Bouchet2015}.
MIRI  imaging will be a factor of 50 more sensitive than {\em Spitzer's}  IRAC and MIPS instruments with a seven fold improvement in resolution; the field-of-view for the MIRI imager (MIRIM) is 74'' $\times$  113'', with a plate scale of 0.11 arcsec/pixel and a full-width half maximum of $\sim$0.7'' \citep{Bouchet2015}. MIRIM provides continuous mid-IR wavelength coverage for broad-band imaging in nine filters from 5.6--25.5 $\mu$m; these filters have been optimized for the detection of the most astronomically relevant molecular and dust species. For instance, the F1130W and F1000W filters are sensitive to PAH's and silicate dust respectively.  

The MIRI imager is significantly more sensitive than the Medium Resolution Spectrograph (MRS), with a larger field-of-view, which means that studies of stellar population on a galactic scale can be done in a reasonable time. The down side is that it is more difficult to correctly identify the nature of the observed object. 
In this paper we will evaluate potential photometric classifications with {\em JWST}/MIRI using the stellar populations of the Large Magellanic Cloud (LMC) as a template. In Section \ref{sec:method} we describe our method for calculating the MIRI synthetic photometry. Section \ref{sec:results} presents the results of these calculations and shows some illustrative MIRI colour-magnitude diagrams for the stellar populations of the Magellanic Clouds. Finally in Section \ref{sec:LocalVol} we discuss how to apply this study to more distant galaxies. 
   

\section{Method} \label{sec:method}

The nine {\em JWST}/MIRI broadband imaging filters will generate a wealth of photometric data for mid-IR stellar populations. In the following section we use the MIRI filter curves from \cite{Glasse2015} to develop colour selection criteria that can isolate and differentiate IR stellar populations.
This analysis will provide guidance for planing {\em JWST} observing programs and determining point-source classifications across a range of metallicities. 

\subsection{Mid-Infrared Spectroscopic Templates}

\subsubsection{Observational data}
\label{sec:AGB_YSO_obs}

In order to predict fluxes and colours through the MIRI filters we have made use of {\em Spitzer}-IRS spectra \citep{Houck2004}, from the SAGE-Spec legacy survey of the LMC \citep{Kemper2010}. The IRS spectra cover a wavelength range from 5.3 to 38 $\mu$m with spectral resolutions, R$=\lambda$/$\Delta\lambda$ $\sim$60--600. The distance to the LMC is well known \citep[see][]{Ngeow2008, Szewczyk2009, Feast2013},  the galaxy's inclination angle is nearly face on ($\sim$30 degrees; \citealt{vanderMarel2001}) and all the stellar populations are at essentially the same distance, thus, the LMC {\em Spitzer}-IRS spectra can be used to predict fluxes and colours for any photometric filter within this range.  As there is no spectral coverage by the {\em Spitzer}-IRS at $\lambda<5.3$ $\mu$m we cannot predict fluxes for the F560W filter.  
For all sources in the LMC we adopt a distance modulus of 18.49 $\pm$ 0.05 \citep{Pietrzynski2013}. 


Over 1000 point sources in the LMC have been observed with the {\em Spitzer} IRS and homogeneously reduced by the SAGE-Spec {\em Spitzer} legacy program \citep{Kemper2010, Woods2011b}, yielding a coherent spectral catlaogue.
The reduced {\em Spitzer} IRS data can be obtained from the NASA/IPAC Infrared Science Archive.\footnote{\url{http://irsa.ipac.caltech.edu/data/SPITZER/SAGE}} These sources cover the range in luminosities and colours parametrized by IRAC, MIPS, and 2MASS magnitudes found in the SAGE-LMC photometric survey \citep{Meixner2006}. They have been classified  in a uniform manner with respect to their base properties according to their mid- and far-infrared spectral characteristics,  SED shape, pulsation period, and bolometric magnitude by \citet{Woods2011, Ruffle2015} and Jones et al.~(in press)  using a classification flow chart (fig.~3 in \citealt{Ruffle2015}).

The point sources observed by the SAGE-Spec survey have been separated according to their evolutionary stage (young stellar objects, main sequence star, asymptotic giant branch star, post-asymptotic giant branch and planetary nebula), chemistry (Oxygen- or Carbon-rich) and by mass in the case of red supergiants. The Young Stellar Object (YSO) class has been further subdivided on the bases of their spectral features, i.e.~embedded YSOs (YSO-1) have silicate and ice absorption features at 10 $\mu$m and 15 $\mu$m; conversely late-type massive YSOs (YSO-3) have silicate features in emission. As these stars become hotter and excites its environment (they develop a compact \Hii region) the spectra exhibit Polycyclic Aromatic Hydrocarbons (PAHs) emission features and atomic emission lines, indicative of a UV radiation field. The intermediate-mass YSO-4 class corresponds to candidate Herbig AeBe (HAeBe) stars.  The spectral features of YSOs can be used as a proxy for evolutionary stage: evolving from heavily embedded, YSO-1s  to compact \Hii regions.

A limited number of sources classified as YSO-3 may be confirmed as compact \Hii regions with {\em JWST}. This confusion arises because evolved YSOs and (ultra)compact \Hii region have a continuum of similar properties, and the spatial resolution probed by the {\em Spitzer}-IRS in the LMC is at best $\sim 1 pc~(4.2'')$. This makes it challenging to resolve compact \Hii regions and separate the contributions of the YSO from its environment.  Readers interested in evolved-YSOs should therefore consider both the YSO-3 and \Hii categories. 

Several mid-IR spectroscopic studies of these spectral-classes have been published in the literature. \citet{Buchanan2006, Buchanan2009} obtained and classified the spectra of 123 of the 250 most luminous 8 $\mu$m sources in the LMC. Whilst, \citealt{Kraemer2017} classify the bright mid-infrared population of the SMC. Objects specifically targeted were various intermediate-to-high-mass post-main sequence stars or stars which are undergoing significant mass loss.

Evolved stars were the focus of several {\em Spitzer}-IRS programs (PID: 200, 1094, 3591, 3505, 3583, 50147 and 50167); their selection was predominantly based on near- and mid-IR colour classification schemes (e.g.~\citealt{Egan2001} for GTO and cycle 1--3 programs or the SAGE evolved star photometric cuts by \citealt[][for later cycles]{Blum2006}), and in some instances includes variables identified in ground-based programs.
 The spectral properties of the carbon stars observed in the Magellanic Clouds was published by \citet{Sloan2016}; the sample of O-rich AGB stars and RSGs was published by \citep{Jones2012}. A large subset (145) of the dust-producing AGB and RSG spectroscopic sample has independently been characterized by \citet{Groenewegen2009}.

Post-AGB stars and PNe were targeted by programs: 103, 20443, 30788, 50092 and 50338, and their spectra published by \citet{Stanghellini2007, BernardSalas2008, BernardSalas2009, Volk2011, Gielen2011, Matsuura2014, Sloan2014}. The spectra of the PNe were obtained early on in the {\em Spitzer} mission; the majority of these sources had been previously observed with {\em HST}.  Post-AGB stars were only explicitly targeted late on in {\em Spitzers} cryogenic mission. Candidate post-AGB stars were identified using a combination of flux limits and  mid-IR colours to exclude YSOs and supergiants. The post-AGB candidates were then cross-matched with the literature to ensure a clean selection.

There are $\sim$300 YSO and H{\sc ii} spectra observed in the LMC, these are predominantly from program 40650 (PI: Looney) and their spectra published by \citet{Seale2009}. YSO spectroscopic targets were selected from the \citet{Gruendl2009} catalogues of YSO candidates. These catalogues of high-mass YSOs in the LMC, were produced using aperture photometry on the SAGE images.

The LMC spectroscopy available in the {\em Spitzer} archive is biased towards the science goals of the PI and the selection effects of the original programs. Spectroscopic studies performed in cycles 1--3  were skewed towards the brightest sources in the LMC as target selection was limited by the Midcourse Space Experiment (MSX; \citealt{Egan2001}) sensitivity limits (7.5 mag at 8 $\mu$m) or was biased towards categories of object known prior the launch of {\em Spitzer}. The reader is referred to table 2 of \citet{Kemper2010} for a comprehensive description of all the {\em Spitzer} programs which have targeted objects in the LMC with the IRS.

To eliminate biases in the LMC spectroscopic sample, the {\em Spitzer} SAGE-Spec program (PID: 40159) targeted unexplored and underrepresented region of IRAC/MIPS colour-magnitude space for a range of object classes.  Spectroscopic candidates were carefully selected to cover the full range in luminosities and colour found in the SAGE data, whilst simultaneously sampling the key phases of stellar evolution. The sensitivity limit of the SAGE-Spec survey is $[8.0] = 7.78 + 0.98 \times ([8.0]-[24])$ and corresponds to a bolometric magnitude of M$_{\rm bol} < -3.75$. It should also be noted that our spectral catalogue is not very sensitive to low-mass ($<$5 M$_{\odot}$) sources  and in crowded star-forming regions some of the IRS sources may be small unresolved clusters \citep{Ward2017} which would be resolved into multiple components with {\em JWST}.


The {\em Spitzer}-IRS spectra sample the complete range of object classes found in the infrared stellar populations of the Magellanic Clouds. 
Table~\ref{tab:classSummary} gives a summary of the spectral classification groups used by \citet[][]{Woods2011, Ruffle2015} and Jones et al.~(in press).
This spectral inventory can act as an empirical template for analyzing and interpreting future infrared data on stellar populations for more distance galaxies in both the local and high-redshift universe. 


As a check on the flux calibration of the spectra, we compare the IRS spectra to the SAGE photometry at 8.0 and 24 $\mu$m. The absolute flux calibration of the SAGE photometry has a higher fidelity than the {\em Spitzer} IRS spectra and is accurate to 3\% \citep{Rieke2004,Reach2005,Engelbracht2007,Bohlin2011}.
 Spectrophotometry for each source was synthesized from the {\em Spitzer} spectra at the effective central wavelength in each filter and checked against the observed IRAC and MIPS observations at 8 and 24 $\mu$m. About 17\% of the spectral observations are offset from the {\em Spitzer} SAGE photometry by over 10\%. In some cases this is due to stellar variability, as both AGB stars and YSOs can show strong variations in brightness. To alleviate disparities in the flux, we exclude sources which show deviations which cannot be accounted for by pulsations.  These sources are not included in Table~\ref{tab:classSummary}. Any remaining systematic effects are minimal since we are looking at a large population of stars.

\begin{table}
\caption{Classification types used by SAGE-Spec \citep{Woods2011}, which we adopt throughout this paper.}
\label{tab:classSummary}
\setlength{\tabcolsep}{10pt} 
\begin{tabular}{llc}
\hline 
Code	&  Object type				&  Number	\\
\hline                                            
C-AGB	&  Carbon-rich AGB stars		&            145	\\
C-PAGB	&  Carbon-rich post-AGB stars		&  \phantom{1}19	\\
C-PN	&  Carbon-rich planetary nebulae	&  \phantom{1}13	\\
O-AGB	&  Oxygen-rich AGB stars		&  \phantom{1}73	\\
O-PAGB	&  Oxygen-rich post-AGB stars		&  \phantom{1}23	\\
O-PN	&  Oxygen-rich planetary nebulae	&  \phantom{1}27	\\
RSG	&  Red Supergiants			&  \phantom{1}74	\\
STAR	&  Stellar photospheres			&  \phantom{1}30	\\
YSO-1	&  Embedded Young Stellar Objects	&  \phantom{1}53	\\
YSO-2	&  Young Stellar Objects		&  \phantom{1}14	\\
YSO-3 	&  Evolved Young Stellar Objects	&  \phantom{1}77	\\
YSO-4	&  HAeBe Young Stellar Objects		&  \phantom{1}21	\\
HII     &  H\,{\sc ii} regions	                &         134  \\
GAL     &  Galaxy                               &         136  \\
\hline 
\end{tabular}
\end{table}


In addition to stellar sources, we also	incorporate into our sample {\em Spitzer}-IRS spectra for a broad range of galaxy morphologies, including ellipticals, spirals, merging galaxies, blue compact dwarfs, and luminous infrared galaxies. The galaxies in our sample are within $z<0.05$ and their properties have been summarized by \cite{Brown2014}. Low resolution IRS spectra for these sources covering the 5.2--38 $\mu$m wavelength range were obtained from version seven of the Cornell Atlas of {\em Spitzer} IRS Sources \citep{Lebouteiller2011}, using the tapered column extraction that is optimized for extended sources. The inclusion of these galaxies  (denoted as {\tt GAL}) in our sample allows us to determine the level of confusion between the stellar population of the host galaxy and unresolved background contaminating point sources. 


\subsubsection{Data from radiative transfer models}
\label{sec:AGB_YSO_models}

To complement the observed {\em Spitzer}-IRS spectra we have also obtained model data from the Grid of Red supergiant and Asymptotic giant branch star ModelS ({\sc grams}; \citealt{Sargent2011, Srinivasan2011}), and the grid of YSO models developed by \citet{Robitaille2006}. These model grids have the advantage over observed sources in regard to lack of noise in the data and the fundamental knowledge about their stellar properties. This allows us to trace the evolution of a source using physical quantities in colour-magnitude space. It also enables us to probe different mass regimes to what the {\em Spitzer}-IRS sample is sensitive too. 
 However, the models suffer from some limitations; they do not account for molecular nuances and the complex dust emission from various species present in the spectra, for instance the YSO models do not include contributions from PAH emission or atomic emission lines. They also do not include multiple stellar sources, which can influence the near- and mid-IR emission. As both the observed spectra and the models have their strengths an weaknesses we use both methods to investigate MIRI colours.


 The {\sc grams} models of  carbon- and oxygen-rich AGB stars and RSGs samples a large range of stellar and spherical dust shell parameters relevant to evolved stars which are undergoing mass loss. The {\sc grams} models were produced using the {\sc 2Dust} radiative transfer code \citep{Ueta2003} and were constructed around model photospheres computed by \citet{Kucinskas2005}  and \citet{Aringer2009} for the M and C stars, respectively. The oxygen-rich models were computed by \citet{Sargent2011} using oxygen-deficient silicate grains (i.e.~silicates that have not reached their stochiometric proportions) from \citet{Ossenkopf1992}, while the carbonaceous dust \citep{Srinivasan2011}  is composed of a mixture of 90\% amorphous carbon
 \citep{Zubko1996} and 10\% silicon carbide (SiC; \citealt{Pegourie1988}). The {\sc grams} models do not distinguish O-rich AGB stars from RSGs.
For both the carbon and oxygen rich grids, we select models with a stellar effective temperature of  T$_{\rm eff}$ = 2100--4700 K, a dust shell inner radius of R$_{\rm in}$ = 3 or 7 R$_{\rm star}$ and log(g) = -0.5. These parameters were chosen as they are representative of the range of values expected for AGB stars, and together with mass-loss they have greatest influence upon the model output.


For the YSOs we use the precomputed 2D radiative transfer model grid developed by \citet{Robitaille2006}. These models cover a range of stellar masses from 0.1--50 M$_{\odot}$ and assume a young central source (0.001--10 Myr) with a rotationally flattened infalling envelope, bipolar cavities and a flared accretion disk. Each model SED is computed at 10 inclination angles. 
 The dust in the YSO models is represented by a mixture of astronomical silicates and graphite at solar abundances from the optical constants of \citet{Laor1993}. The grain size distribution varies with location in the disk and envelope. 
The models are divided into three stages, ranging from the early envelope infall to when the young star is dispersing its protoplanetary disk.

Stage I YSOs have substantial envelopes, and the central source accretes dust and gas at rates of  $\dot{M}_{\rm env}/{M_{*}}> 10^{-6}\,{M_{\odot }}$ yr$^{-1}$.
Stage II YSOs have a dispersed protostellar envelope and optically thick disks $\dot{M}_{\rm env}/{M_{*} }< 10^{-6}\,{M_{\odot }}$ yr$^{-1}$ and $M_{\rm {disk}}/M_{\star }>10^{-6}$.
Finally, stage III YSOs have optically thin disks $\dot{M}_{\rm env}/{M_{*} }<10^{-6}\,{M_{\odot }}$ yr$^{-1}$ and $M_{\rm {disk}}/M_{\star }<10^{-6}$.

 The \citet{Woods2011} YSO-1 and YSO-2 objects are synonymous with the Stage I sources from the \citet{Robitaille2006} models; whilst the YSO-3 and YSO-4 class is the analogue of Stage II objects from the high and intermediate mass regimes, respectively.

As the relative importance of the various parameters in the \citet{Robitaille2006} models are dependant on the given evolutionary stage, we do not limit a parameters range to follow a particular star throughout its evolution, instead we focus on YSOs in three stellar mass regimes: 0.2, 2.0 and 20 $\pm 2.5\%$ $M_{\odot }$ at a disk inclination of 48.5$^{\circ}$. These values are representative of the minimum YSO mass a {\em JWST} program may detect in a reasonable integration time for the Milky Way, the Magellanic Clouds and more distant local group galaxies.

\subsection{Predicted MIRI Photometry}

In order to calculate predicted MIRI broadband flux densities and colours for mid-IR stellar populations we use the same convention outlined by earlier infrared space telescopes such as {\em IRAS, ISO, Spitzer} and {\em WISE} \citep[e.g.][]{Beichman1988, Blommaert2003, Reach2005, Wright2010}. Here photometry for the desired photometric band $f_{\nu}^{\rm{MIRI}}(\lambda_{\rm{eff}})$ is synthesised from a convolution of the template source spectra $F_{\nu}$ with the MIRI filter functions from \cite{Glasse2015}.

The MIRI relative system response function $R_{\gamma}$, defined as the fraction of detected electrons per photon crossing the focal plane of the MIRI Imager were created by \cite{Glasse2015} from two independent measurements. The response function for the MIRI filters are in units of electrons per photon. This can be converted to a photon-counting response function $R$ in units of electrons per unit energy, by re-normalizing the product of $\lambda R_{\gamma}$ \citep[e.g.~][]{Bessell2000}. 


Fluxes in the MIRI wavebands were calculated by integrating the {\em Spitzer}/model spectra ($F_{\nu}$) of each source over the MIRI spectral response according to the following equations:
\begin{equation}
\label{k_short}
f_{\nu}^{\rm{MIRI}}(\lambda_{\rm{eff}}) = F_{\nu}(\lambda_{\rm{eff}}) \times K ,
\end{equation}
where $\lambda_{\rm{eff}}$ is the effective wavelength of the filter, $F_{\nu}(\lambda_{\rm{eff}})$ is the flux density of the input {\em Spitzer} spectrum at $\lambda_{\rm{eff}}$ and $f_{\nu}^{\rm{MIRI}}(\lambda_{\rm{eff}})$ is the monochromatic flux density that would be given if the source was observed by the MIRI imager. 
The dimensionless quantity $K$ is defined in terms of frequency units by
\begin{equation}
\label{k}
K = {\int (F_{\nu}/F_{\nu_o})(\nu/\nu_{\rm{eff}})^{-1}~R_{\gamma}~d\nu \over {\int(\nu/\nu_{\rm{eff}})^{-2}~R_{\gamma}~d\nu}}~, 
\end{equation}

where $F_{\nu_o}$ is a reference spectrum, this is assumed to be a $\nu F_\nu$ = constant flux spectrum \citep[see e.g.][]{Blommaert2003, Reach2005, Hora2008, Bohlin2011}.

Alternatively, the flux can be calculated in terms of the photon weighted mean flux over the bandpass.  This effective flux is defined in frequency units as
\begin{equation}
\langle F_{\nu}\rangle={\int F_\nu~\nu^{-1}~R_{\gamma}~d\nu \over \int \nu^{-1}~R_{\gamma}~d\nu}.  
\label{favnu}
\end{equation}
This method is traditionally used by {\em HST} and optical telescopes; it does not involve colour corrections or effective wavelengths, prevalent in flux calculations for IR space-based telescopes. 

The choice of methodology between equations~\ref{k_short} and \ref{k}, and equation~\ref{favnu} can return a flux for the MIRI filters ($f_{\nu}^{\rm{MIRI}}$) which differs by a few percent (typically $\lesssim 3 \%$). However the divergence can be more severe if the shape of source spectrum significantly differs from that of the reference spectrum. By convention the Jansky systems assumes a  $\nu F_\nu$ = constant  reference system.  



When calculating uncertainties associated with the MIRI fluxes; we first consider the flux uncertainties of the observed {\em Spitzer} IRS sample and the absolute flux calibration errors. The formal uncertainties in the mean ($\sigma /\sqrt{N}$) and systematic uncertainties from the absolute flux calibration \citep{Sloan2015a}; the presence of other sources in the slit; and pointing errors, which introduce discontinuities between the segments, need to be assessed. The flux uncertainties and calibration are described in detail by \citet[][]{Woods2011b, Lebouteiller2011}, and are summarized in the appendix.

To estimate the total uncertainty of a flux we use the quadratic sum of these measured and systematic errors. For the {\sc grams} and YSO models we assume a total uncertainty of 5\%, to reflect the uncertainty in the stellar atmosphere models.


The final uncertainties in the MIRI fluxes were computed by propagating the uncertainties from the spectra and models.

\begin{figure}
\centering
\includegraphics[trim=1.0cm 0cm 0cm 0cm, width=3.5in]{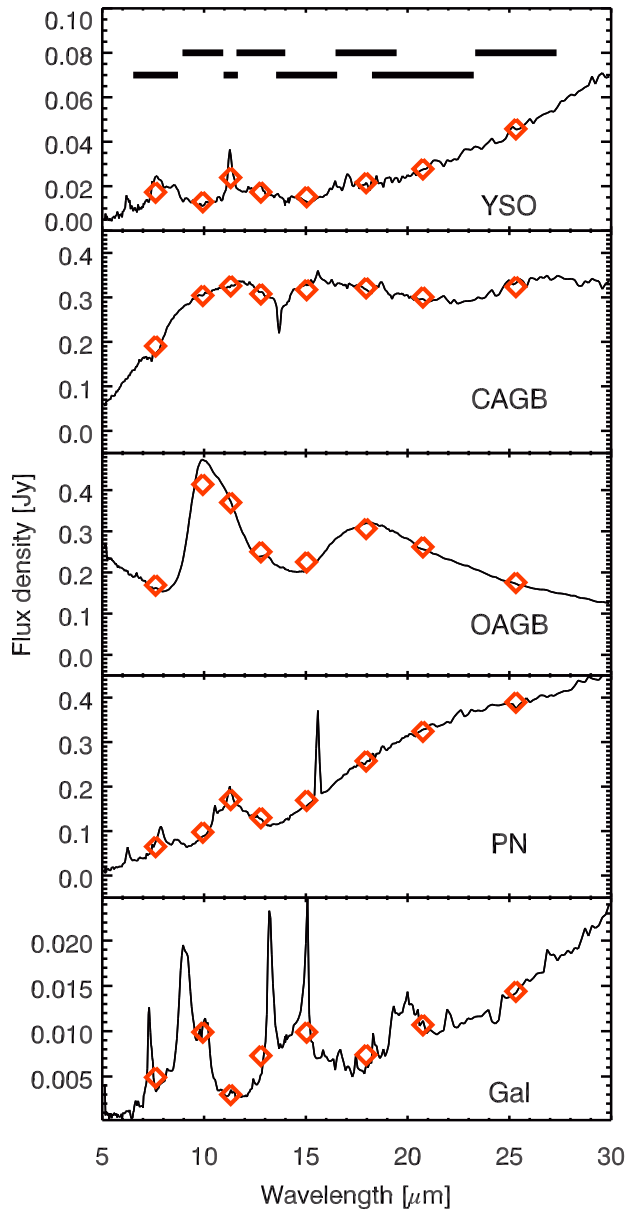}
\caption{{\em Spitzer}-IRS spectra of a young stellar object, a carbon-rich AGB star, an oxygen-rich AGB star, a planetary nebulae, and the rest-frame spectrum of a star-forming galaxy. The resulting fluxes from integrating the spectra over the MIRI bandpasses are shown in red. The thick black lines in the top pannel shows the wavelength coverage of the MIRI bandpasses.} 
\label{fig:typicalSpec}
\end{figure}

In Figure~\ref{fig:typicalSpec} we present the {\em Spitzer}-IRS spectra of a representative subsample of evolved stars, YSOs and galaxies and the MIRI broadband flux densities these sources would have if they were observed with {\em JWST}. These sources are representative of their stellar class and can be characterized by their spectral appearance in the mid-IR.

\subsection{Effective wavelength}

As noted earlier the flux density of a source is evaluated at the effective wavelength of the filter. The optimal choice for the effective wavelength is the filter wavelength which is least sensitive to the spectral shape of the source i.e.~the weighted average wavelength, defined as
\begin{equation}
\lambda_{\rm{eff}} = {\int\lambda^2 R_{\gamma}~d\lambda\over{\int \lambda R_{\gamma}~d\lambda}}~.
\label{eqn:effwav}
\end{equation} 
This is related to the effective-frequency via $\nu_{\rm eff} =c/\lambda_{\rm eff} $.
Table~\ref{tab:Zpoints} gives the effective wavelengths calculated using the current spectral response curves ($R_{\gamma}$) for the MIRI filters \citep{Glasse2015}.


\subsection{Magnitudes}

We estimate zero-magnitude flux densities for the MIRI bands to determine magnitudes of our template sources from their predicted flux densities. We define the magnitude system such that 
\begin{equation}
[M_i] = -2.5{\rm log_{10}}(f_\nu^{{\rm MIRI}, i}/f_{i, {\rm zero}}). 
\label{eq:Mag}
\end{equation} 

The zero-points magnitudes are estimated relative to an `ideal' Vega photospheric reference spectrum, and were determined by integrating a single-temperature Kurucz model spectrum of Vega \citep{Cohen1992} over the MIRI bands, using the method outlined in equations~\ref{k_short} and \ref{k}. This ideal Vega spectrum had a uniform effective temperature across the stellar surface and no infared excess. 

By using the Vega magnitude system we eliminate the dependence on spectral shape, this follows the methodology outlined by \cite{Reach2005} to define the {\em Spitzer}/IRAC Magnitude System and should be comparable to other magnitudes relative to Vega in optical and infrared systems. Table~\ref{tab:Zpoints} gives the resulting zero-magnitude flux densities for the MIRI channels. These zero points do not account for in-orbit telescope performance and will need to be corrected once the total system throughput is verified post-launch.

\begin{figure}
\centering
\includegraphics[trim=1.0cm 0cm 0cm 0cm, width=3.5in]{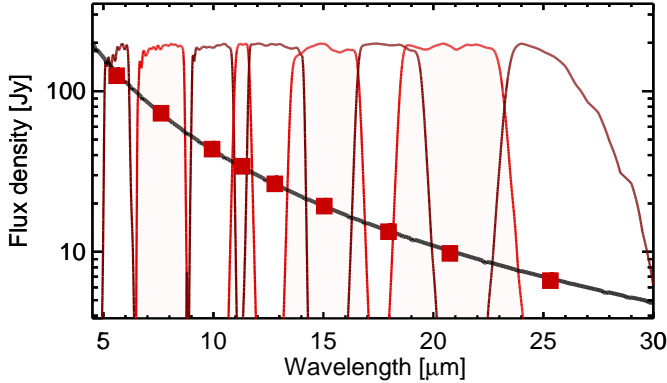}
\caption{An `ideal' Kurucz model spectrum of Vega, together with the relative spectra responses of the MIRI channels (in electrons per photon) normalized to unity. The red squares show the resulting zero-point fluxes.} 
\label{fig:VEGAzoints}
\end{figure}

\begin{table}
\begin{center}
\caption{Estimated effective wavelengths and zero points for the MIRI filters.}
\label{tab:Zpoints}
\begin{tabular}{lccc}
\hline
\hline
Filter	&	Calc.  $\lambda_{\rm{eff}}$ 	&	 $\Delta\lambda$ 	&	Est. Zero Point	\\
	&	$\mu$m	&	$\mu$m	&	Jy	\\
\hline
F560W	&	$-$	&	1.2	&	116.60	\\
F770W	&	7.62	&	2.2	&	68.10	\\
F1000W	&	9.94	&	2.0	&	40.71	\\
F1130W	&	11.31	&	0.7	&	31.82	\\
F1280W	&	12.79	&	2.4	&	24.76	\\
F1500W	&	15.04	&	3.0	&	17.99	\\
F1800W	&	17.96	&	3.0	&	12.46	\\
F2100W	&	20.75	&	5.0	&	9.13	\\
F2550W	&	25.32	&	4.0	&	6.18	\\
\hline
\end{tabular}
\end{center}
\end{table}

\section{Results} 
\label{sec:results}

We derive MIRI synthetic photometry for over 1000 {\em Spitzer}-IRS spectra with known object types. This photometry is analogous to the flux density (in Janskys) that would be obtained if a source was observed with the MIRI instrument. Table~\ref{tab:FluxResults} presents the MIRI flux density calculated according to equation~\ref{favnu}. The source type and predicted MIRI magnitudes calculated according to equation~\ref{eq:Mag} are given in Table~\ref{tab:MagResults}; all reported magnitudes are in the Vega System. 

From this library of stellar classes, MIRI fluxes and magnitudes presented in Tables~\ref{tab:FluxResults} and \ref{tab:MagResults} it is straightforward to compute the expected fluxes and colours for the stellar populations of more distant galaxies. The flux listed in Table~\ref{tab:FluxResults} scales with $1/(d/50)^2$ where d is in kpc and the magnitudes can simply be adjusted to the required distance modulus from that of the LMC, $M-m=$ 18.49 $\pm$ 0.05 mag \citep{Pietrzynski2013}. 

The types of objects in our sample are representative of the IR stellar populations of Local Group galaxies and includes both YSOs and evolved stars. One of the major contaminants of stellar populations studies are unresolved background galaxies. In order to explore the background galaxy population we group all the spectra of individual galaxies into one class (GAL) and do not distinguish between galaxies based on their morphological type. 
It is important to note that the flux of the background galaxies should not be scaled with distance. 

These results can be used to guide the choice of filter with which to observe specific object classes with {\em JWST} and to identify key populations in colour-colour and colour magnitude diagrams once {\em JWST} observations become available. The colour is independent of distance.

\begin{table*}
\begin{center}
\caption{MIRI fluxes (mJy) for our sample, calculated according to equation~\ref{favnu}.}
\label{tab:FluxResults}
\begin{tabular}{lcccccccc}
\hline
\hline
Class & [F770W] & [F1000W] & [F1130W] &	[F1280W] & [F1500W] &	[F1800W] &	[F2100W] &	[F2550W] \\
\hline
GAL	&	4.7	$\pm$	0.3	&	10.0	$\pm$	0.4	&	3.0	$\pm$	0.3	&	7.0	$\pm$	0.4	&	9.8	$\pm$	0.6	&	7.2	$\pm$	0.8	&	10.5	$\pm$	0.5	&	14.4	$\pm$	1.2	\\
C-AGB	&	26.4	$\pm$	1.0	&	27.6	$\pm$	0.9	&	28.9	$\pm$	0.8	&	18.8	$\pm$	1.2	&	14.1	$\pm$	0.7	&	12.3	$\pm$	0.9	&	10.5	$\pm$	0.5	&	8.6	$\pm$	0.5	\\
RSG	&	24.3	$\pm$	1.3	&	18.3	$\pm$	0.7	&	16.5	$\pm$	0.7	&	12.9	$\pm$	1.1	&	10.2	$\pm$	0.7	&	10.1	$\pm$	0.6	&	8.7	$\pm$	0.4	&	7.0	$\pm$	0.5	\\
YSO-2	&	56.6	$\pm$	2.1	&	45.7	$\pm$	1.9	&	59.5	$\pm$	2.3	&	67.8	$\pm$	2.3	&	69.5	$\pm$	2.8	&	72.5	$\pm$	2.9	&	77.5	$\pm$	2.7	&	97.6	$\pm$	5.7	\\
O-AGB	&	22.4	$\pm$	1.2	&	55.4	$\pm$	0.9	&	51.1	$\pm$	0.8	&	33.1	$\pm$	1.1	&	18.6	$\pm$	0.6	&	15.2	$\pm$	0.7	&	12.7	$\pm$	0.5	&	8.9	$\pm$	0.4	\\
\hline
\end{tabular}
\tablecomments{Only a portion of this table is shown here to demonstrate its form and content. A machine-readable version of the full table is available online. A null value of -99.99 indicates that no long-low {\em Spitzer}-IRS data is available for that source.}
\end{center}
\end{table*}

\begin{table*}
\begin{center}
\caption{Estimated MIRI Magnitudes and uncertanties for our LMC and background galaxy sample.}
\label{tab:MagResults}
\begin{tabular}{lcccccccc}
\hline
\hline
 Class & [F770W] & [F1000W] & [F1130W] &	[F1280W] & [F1500W] &	[F1800W] &	[F2100W] &	[F2550W] \\
\hline
GAL   & 10.37 $\pm$ 0.06 & 9.05 $\pm$ 0.04 & 10.05 $\pm$ 0.09 & 8.82 $\pm$ 0.05 & 8.15 $\pm$ 0.06 & 8.07 $\pm$ 0.11 & 7.35 $\pm$ 0.06 & 6.60 $\pm$ 0.04 \\
C-AGB & 8.52 $\pm$ 0.04 & 7.92 $\pm$ 0.04 & 7.60 $\pm$ 0.03 & 7.79 $\pm$ 0.07 & 7.74 $\pm$ 0.05 & 7.51 $\pm$ 0.07 & 7.35 $\pm$ 0.04 & 7.16 $\pm$ 0.06 \\ 
RSG   & 8.62 $\pm$ 0.06 & 8.37 $\pm$ 0.04 & 8.21 $\pm$ 0.05 & 8.20 $\pm$ 0.09 & 8.10 $\pm$ 0.07 & 7.72 $\pm$ 0.06 & 7.56 $\pm$ 0.05 & 7.38 $\pm$ 0.09  \\
YSO2  & 7.68 $\pm$ 0.04 & 7.37 $\pm$ 0.04 & 6.81 $\pm$ 0.04 & 6.39 $\pm$ 0.04 & 6.00 $\pm$ 0.04 & 5.58 $\pm$ 0.04 & 5.18 $\pm$ 0.04 & 4.52 $\pm$ 0.04  \\
O-AGB & 8.69 $\pm$ 0.05 & 7.16 $\pm$ 0.04 & 6.98 $\pm$ 0.04 & 7.18 $\pm$ 0.04 & 7.45 $\pm$ 0.04 & 7.28 $\pm$ 0.04 & 7.15 $\pm$ 0.04 & 7.11 $\pm$ 0.07  \\
\hline
\end{tabular}
\tablecomments{Only a portion of this table is shown here to demonstrate its form and content. A machine-readable version of the full table is available online. A null value of -99.99 indicates that no long-low {\em Spitzer}-IRS data is available for that source.}
\end{center}
\end{table*}


\subsection{MIRI Colours} \label{sec:MIRIcolours}

\begin{figure*}
\centering
\includegraphics[width=72mm]{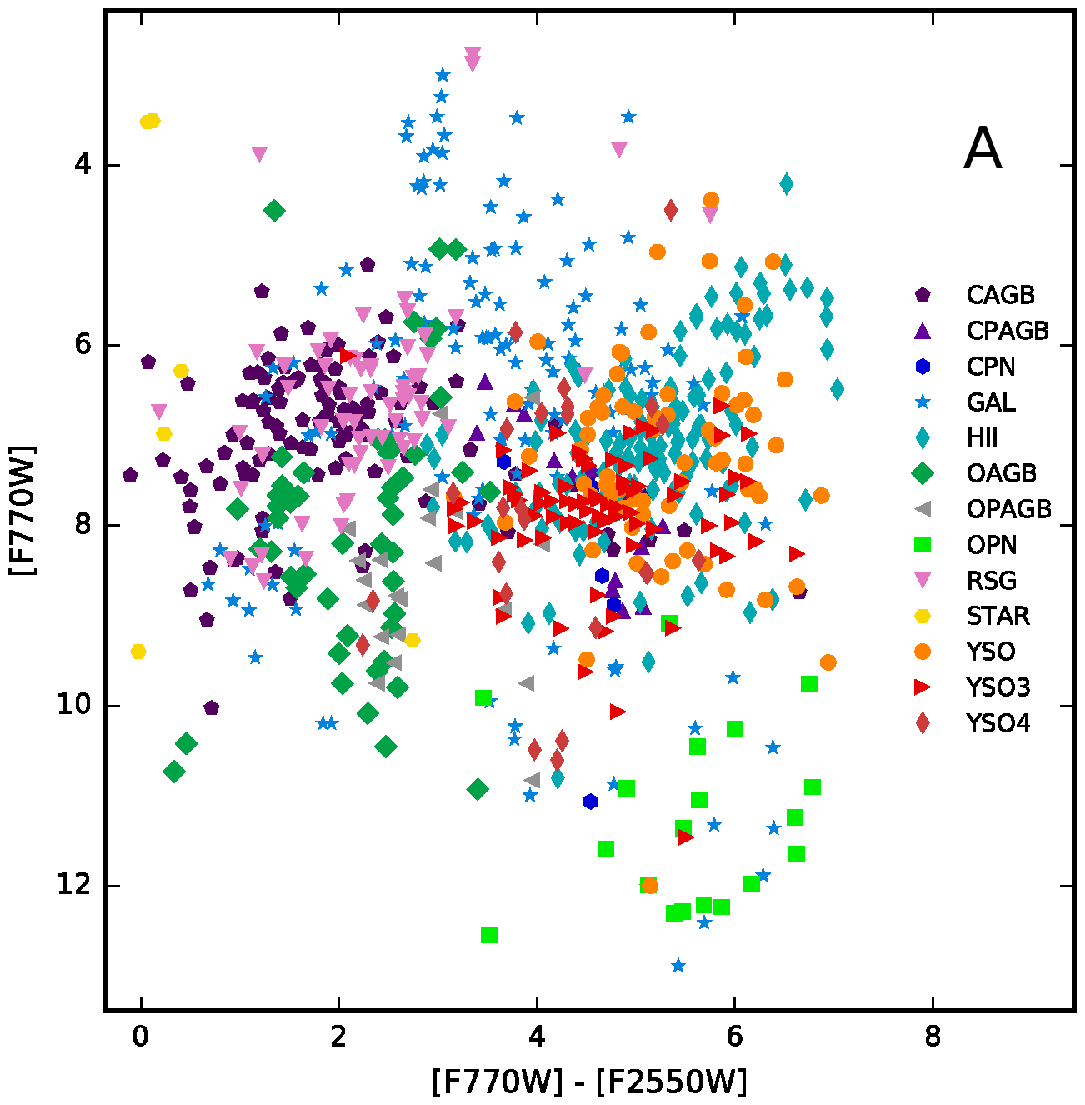}
\includegraphics[width=72mm]{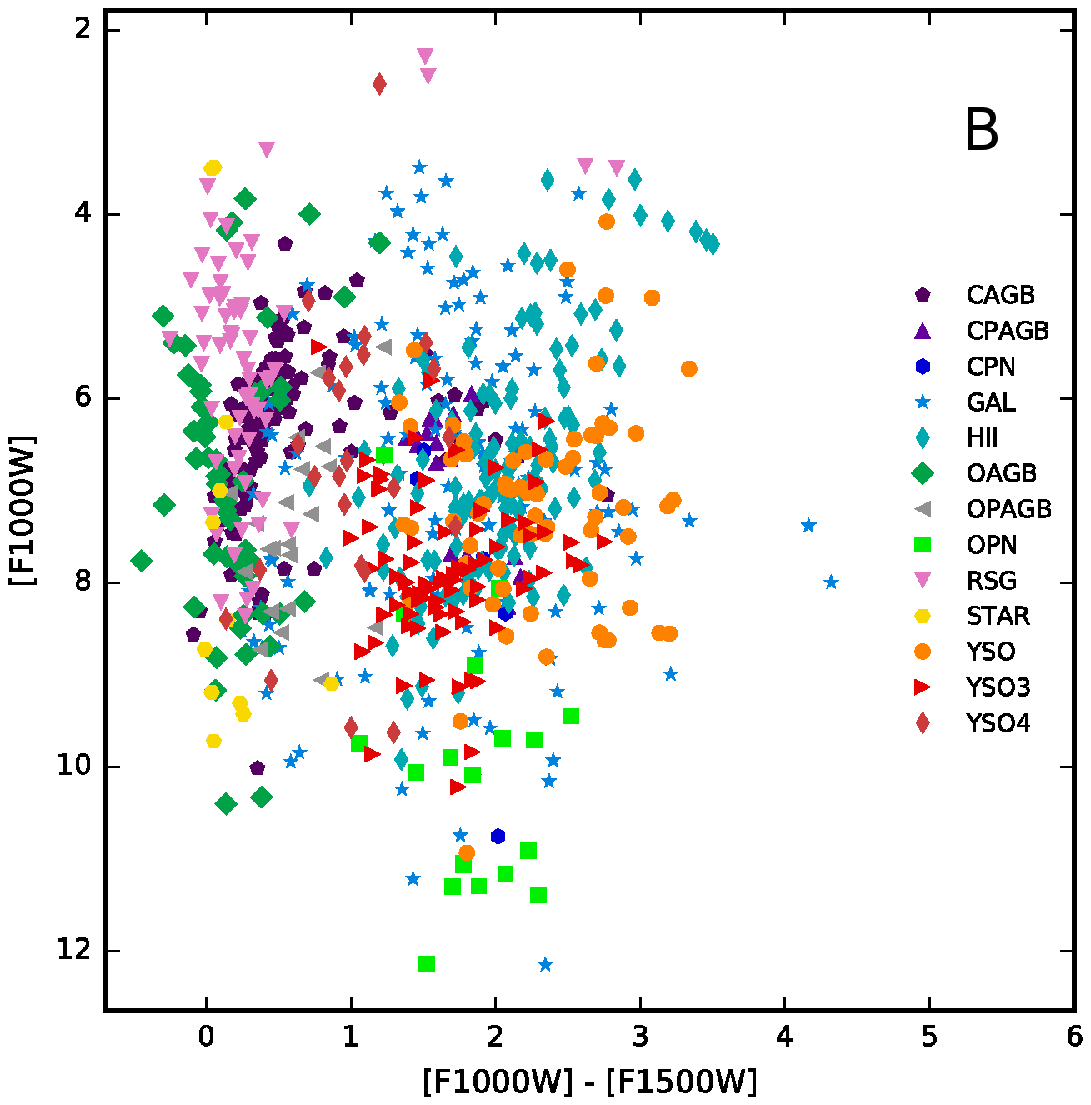}
\includegraphics[width=72mm]{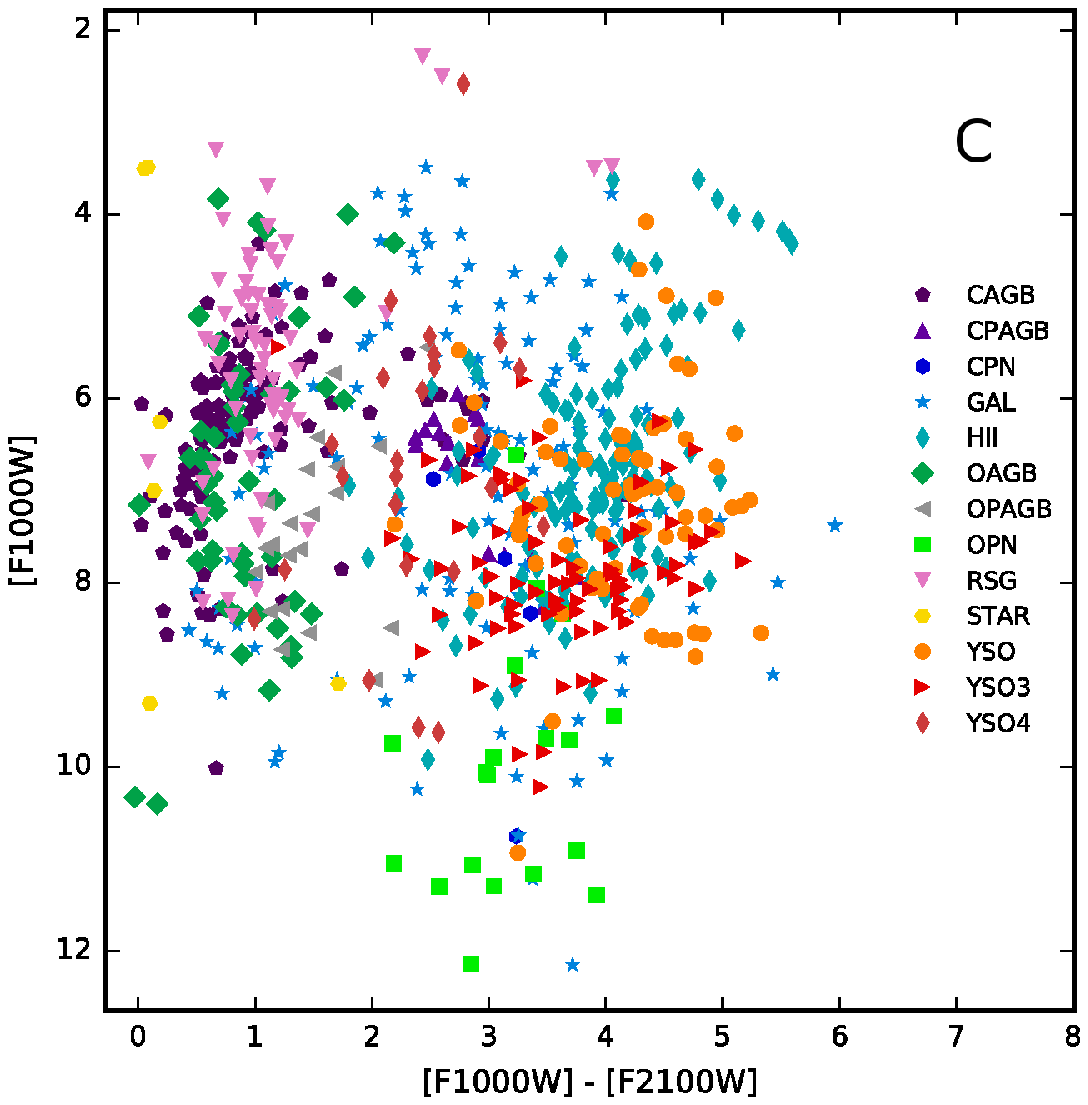}
\includegraphics[width=72mm]{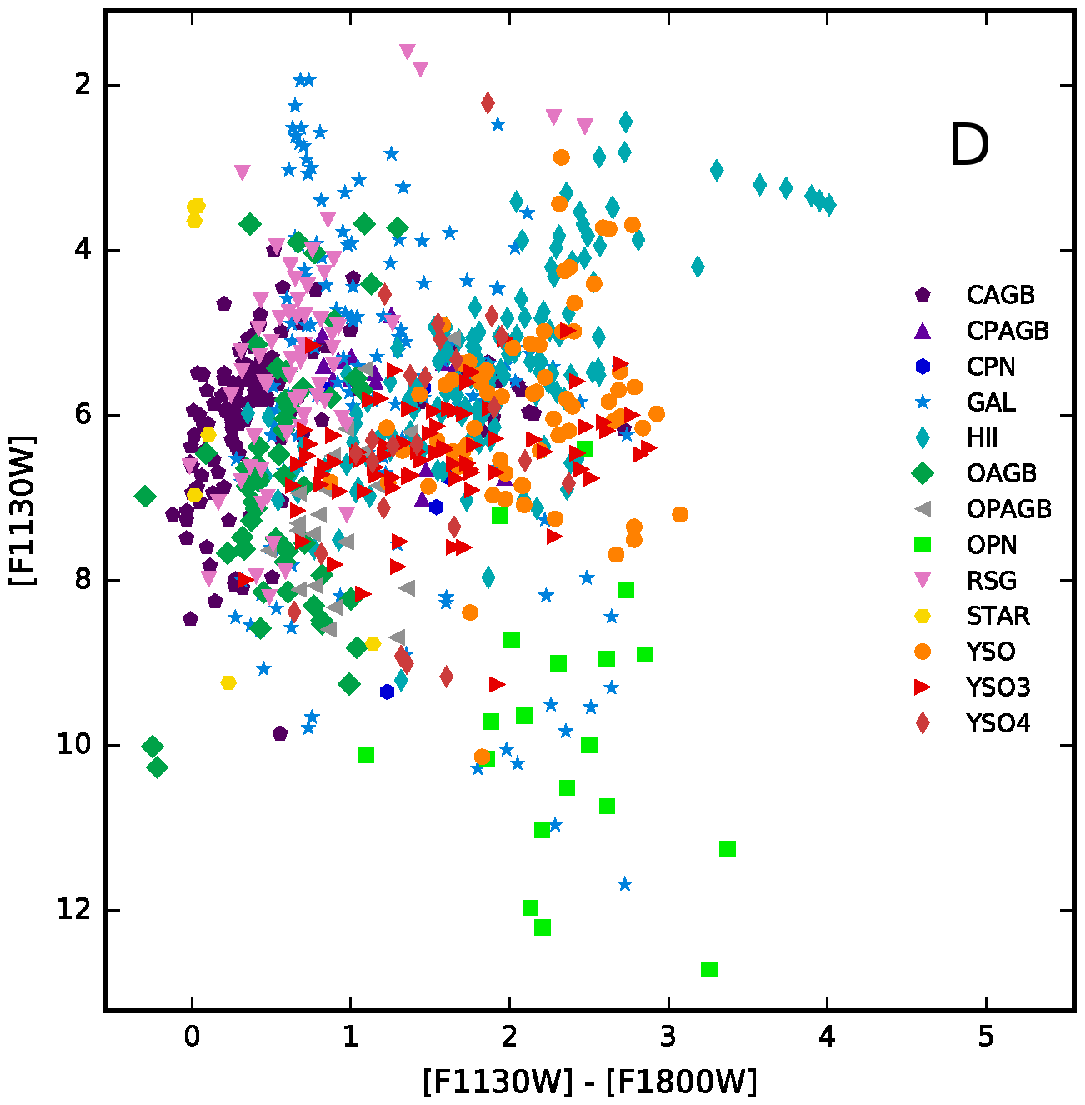}
\includegraphics[width=72mm]{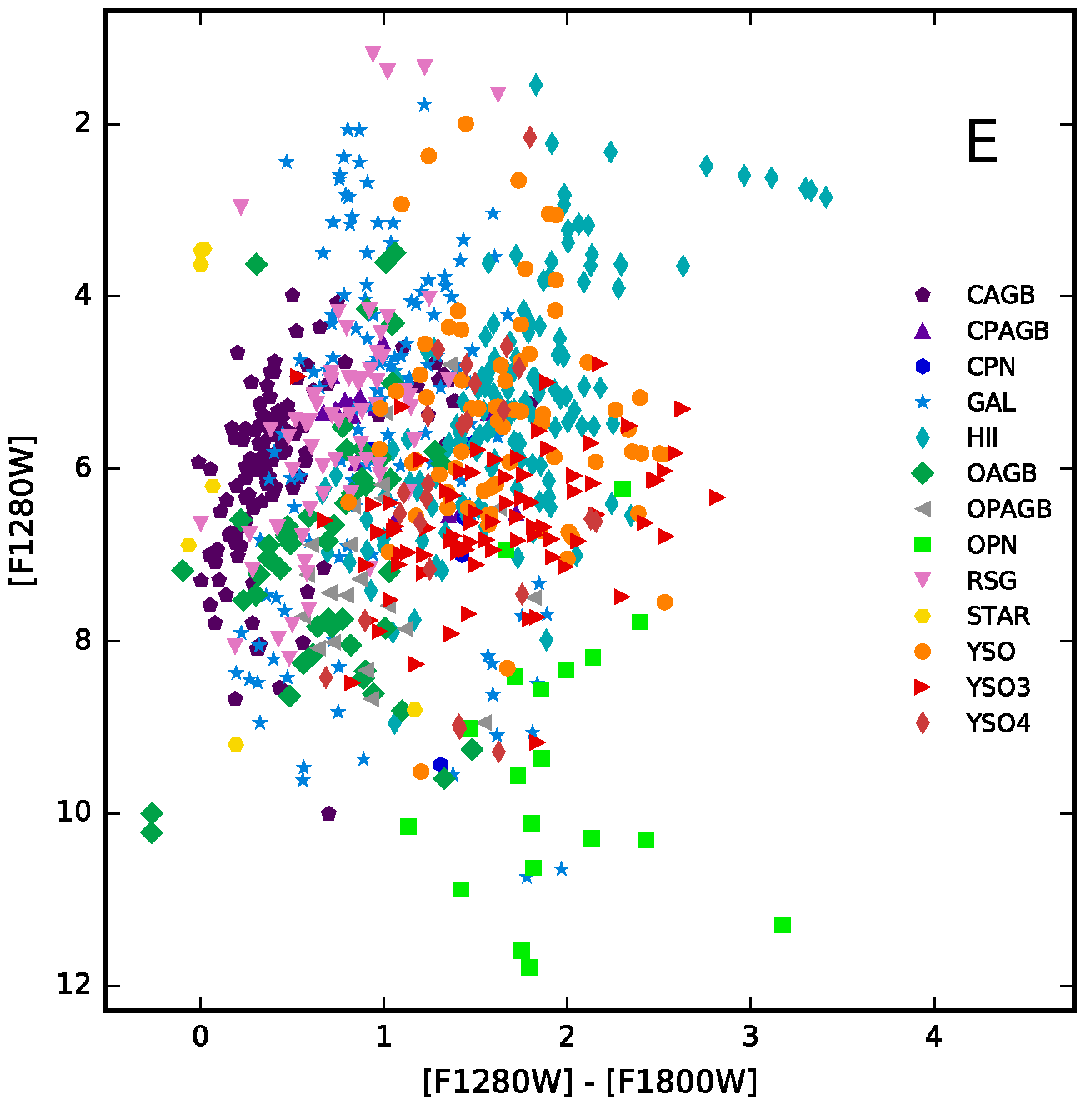}
\includegraphics[width=72mm]{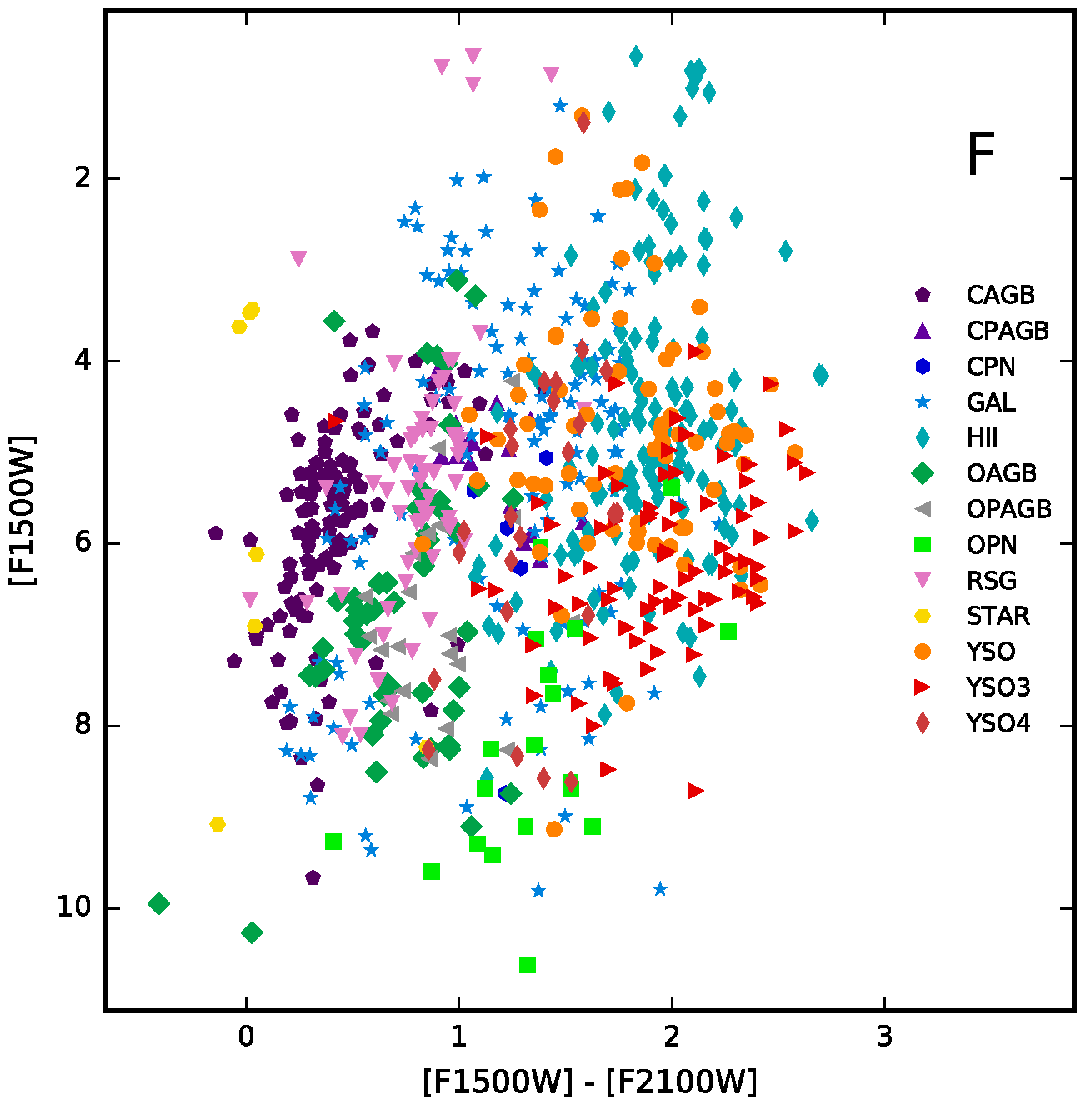}
\caption{Example JWST/MIRI CMDs for over 700 point sources in the LMC. The source are classified by evolutionary stage (e.g.~Young Stellar Objects, asymptotic giant branch, post-asymptotic giant branch and planetary nebula) and by chemistry (Oxygen or Carbon rich).  Each colored symbol shows a different population of sources as indicated in the legend; see Table~\ref{tab:classSummary} for the class definitions.  For clarity the YSO-1 and YSO-2 subcategories are grouped together as YSO. These colors effectively differentiate evolved stars from YSOs and separates carbon stars from oxygen rich stars: this is critical for dust-production rate estimates and dust evolution models.}
\label{fig:MIRICMD}
\end{figure*}

We present example colour-magnitude (CMD) and colour-colour diagrams (CCD) to aid the identification of the various IR stellar and extragalactic populations which will be observed with {\em JWST}.  We estimate that the 3$\sigma$ uncertainty in the colour is up to $\pm$ 0.6 mag.
By employing different filter combinations it is possible to isolate specific populations of objects with minimal confusion.
Figures~\ref{fig:MIRICMD} and \ref{fig:MIRICCD} show example MIRI CMD and CCDs constructed from {\em Spitzer}-IRS spectra with the regions occupied by various types of objects illustrated. Whilst Figures~\ref{fig:MIRICMD_GRAMS} and \ref{fig:YSOCMD} show the results from the evolved star and YSO models.\footnote{Due to the large number of {\em JWST} bands and filter combinations available, we do not show all possible CMD and CCDs, however the interested reader can use the results from Tables~\ref{tab:FluxResults} and \ref{tab:MagResults} to generate CMD and CCDs for any potential filter combinations and science goals. This can be accomplished by either by selecting objects of interest e.g.~just the AGB stars or the population as a whole and then scaling the fluxes or magnitudes to the required distance.}
The CMDs and CCDs (Figures~\ref{fig:MIRICMD} and \ref{fig:MIRICCD}) from the spectral sample, provide an accurate representation of the brightest IR population one may expect in a galaxy (including the variations due to various dust compositions).  The model CMDs (Figures~\ref{fig:MIRICMD_GRAMS} and \ref{fig:YSOCMD}) enable us to assess low-mass sources and rare sources classes which are not evident in the observed sample, and hence provide completeness.
Below we describe the most favourable MIRI filters to identify specific populations of objects with minimal confusion.


\begin{figure}
\centering
\includegraphics[width=72mm]{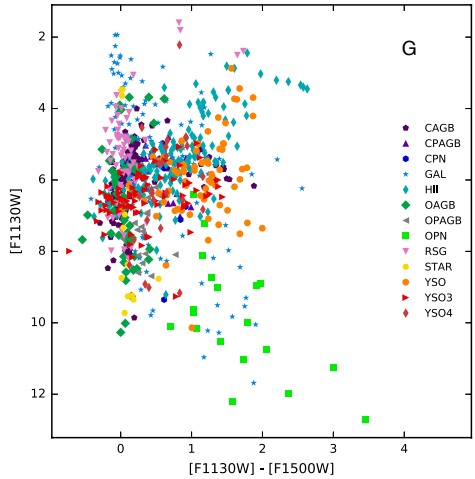}
\caption{Fig.~\ref{fig:MIRICMD} continued.}
\end{figure}

\begin{figure*}
\centering
\includegraphics[width=80mm]{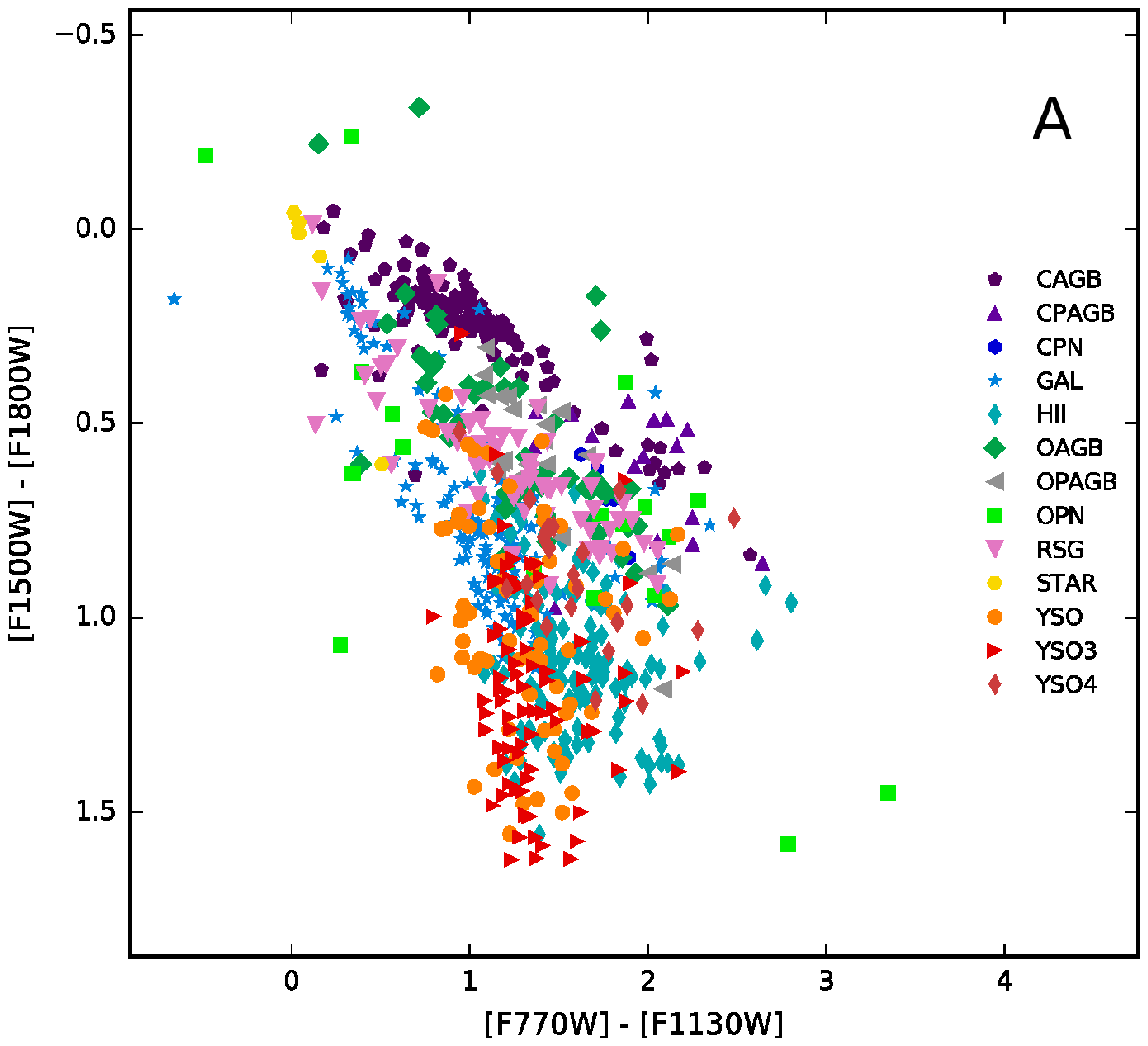}
\includegraphics[width=80mm]{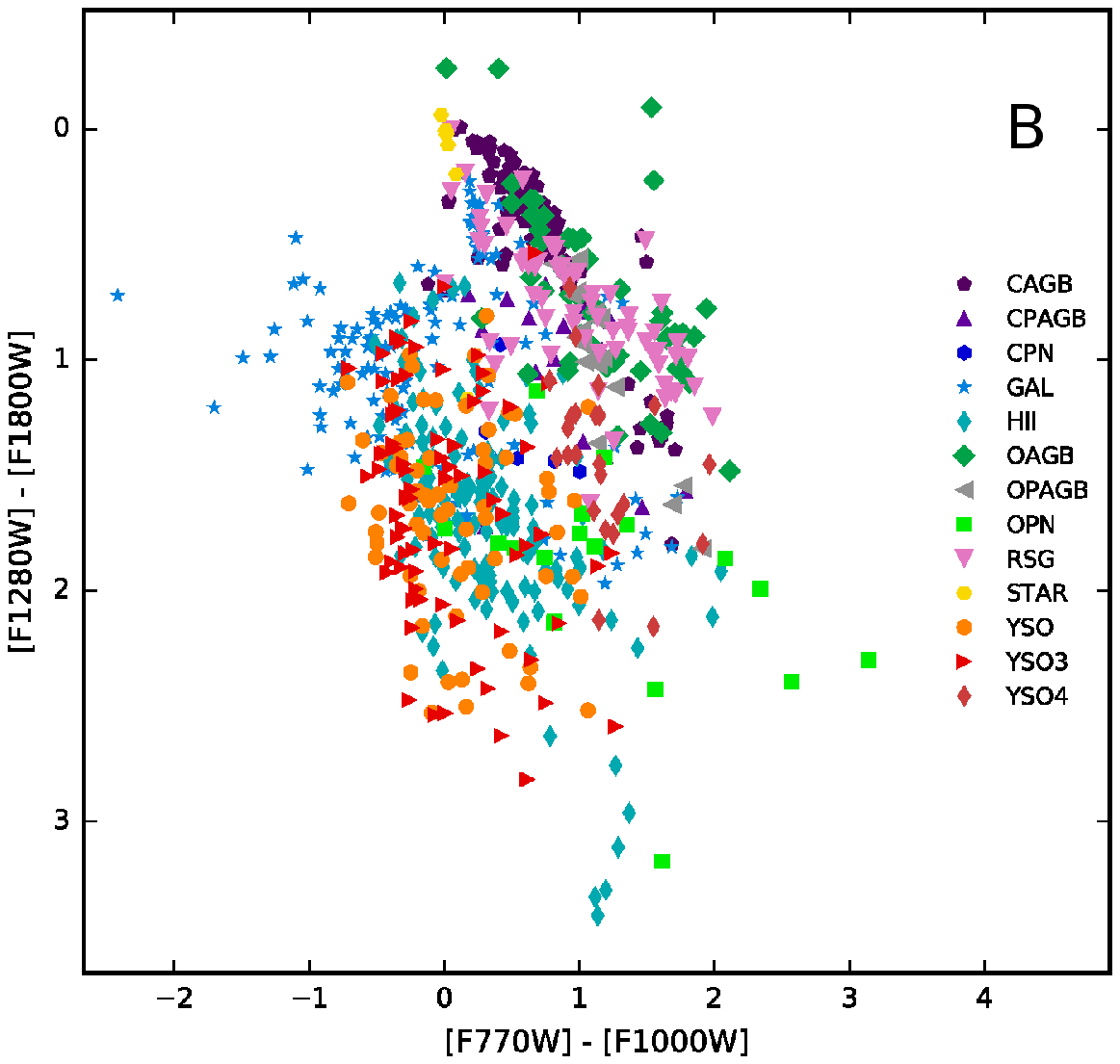}
\includegraphics[width=80mm]{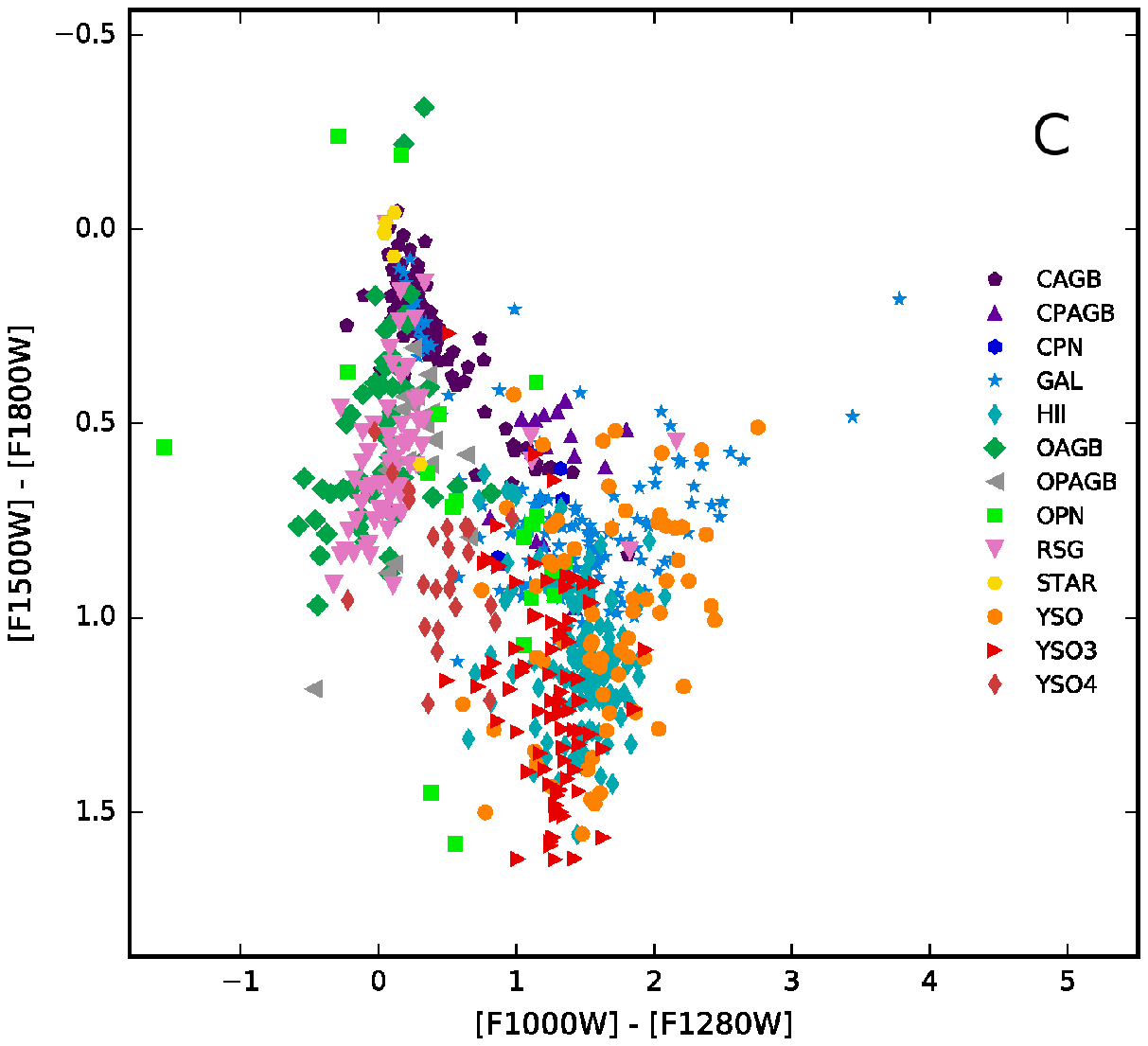}
\includegraphics[width=80mm]{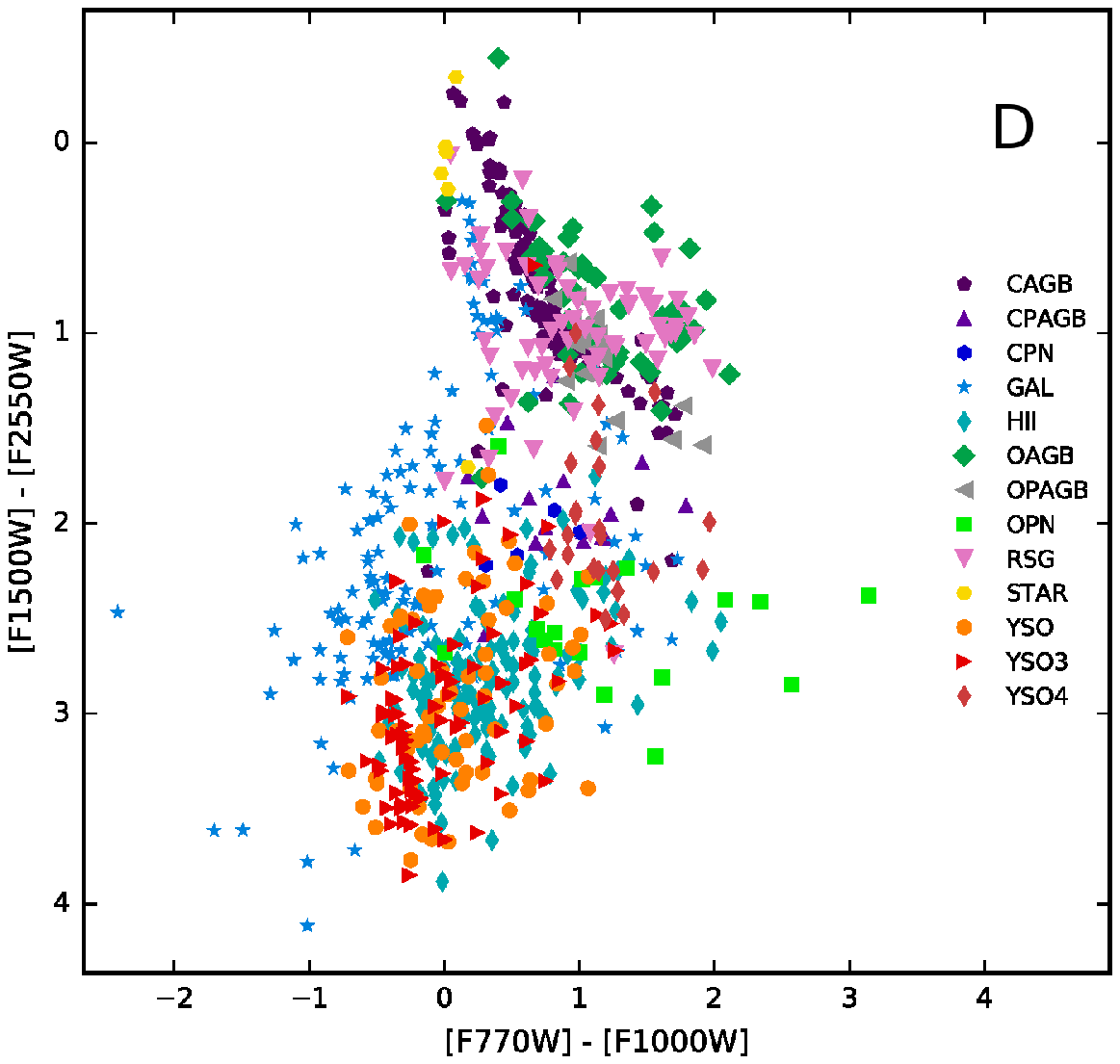}
\caption{Colour-colour diagram showing the locations of interesting classes of objects. These are useful as they are distance independent. Symbols are the same as Fig.~\ref{fig:MIRICMD}.}
\label{fig:MIRICCD}
\end{figure*}


\subsubsection{Main sequence stars}

Stars with a stellar photosphere, but no additional dust or gas features have MIRI colors that fall around zero. The majority of main-sequence and sub-giant stars fall within this class and can only be distinguished on the basis of their short wavelength colours. In MIRI CCDs stars tend to cluster in the upper left side of the diagram (e.g.~Figure~\ref{fig:MIRICCD}, panel A). 

In some instances main sequence OB stars can illuminate and heat patches of dense interstellar medium (ISM) surrounding the star \citep{Sheets2013, Adams2013}. Stars of this nature have colours typical of stellar photospheres at ($\lambda < 8$ $\mu$m) but show a strong infrared excess at longer wavelengths ($\lambda > 20$ $\mu$m) indicative of warm dust associated with cirrus hotspots.

\subsubsection{Evolved stars}

Evolved stars occupy a wide range of MIRI colour-space. Their precise location depends on both the abundance and the chemistry of the dust in their circumstellar envelope. Evolved stars have either oxygen-rich or carbon-rich circumstellar envelopes depending on the C/O ratio in their atmosphere. When C/O $< 1$, the CO molecule ties up the carbon, resulting in oxygen-rich molecules and dust grains. Conversely, when C/O $> 1$ all the oxygen in locked up in CO resulting in carbon-rich molecules and dust. The [F1130W] band centred on the prominent 11.3 $\mu$m PAH  feature and the peak of the Silicon Carbide (SiC) feature, is a powerful diagnostic tool for separating evolved carbon rich sources from oxygen rich sources.  The SiC feature at $\sim$ 11.3 $\mu$m weakens with increasing metallicity, thus this filter may not be as an effective discriminant in metal-rich environments. Carbon-rich post-AGB stars and PNe may show a combination of SiC and PAHs; for these objects the [F1130W] band is a powerful diagnostic.


\begin{figure*}
\centering
\includegraphics[width=80mm]{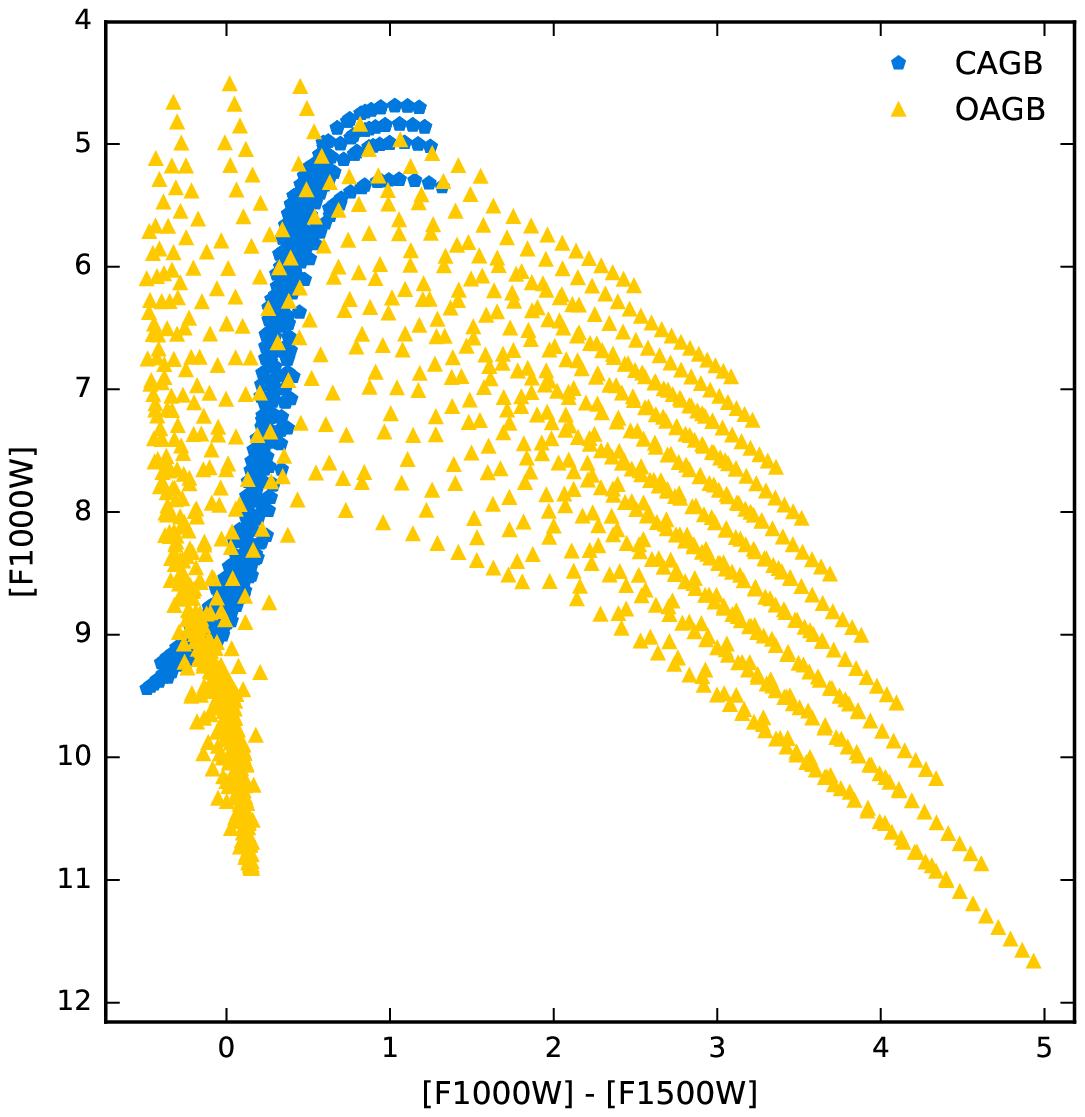}
\includegraphics[width=80mm]{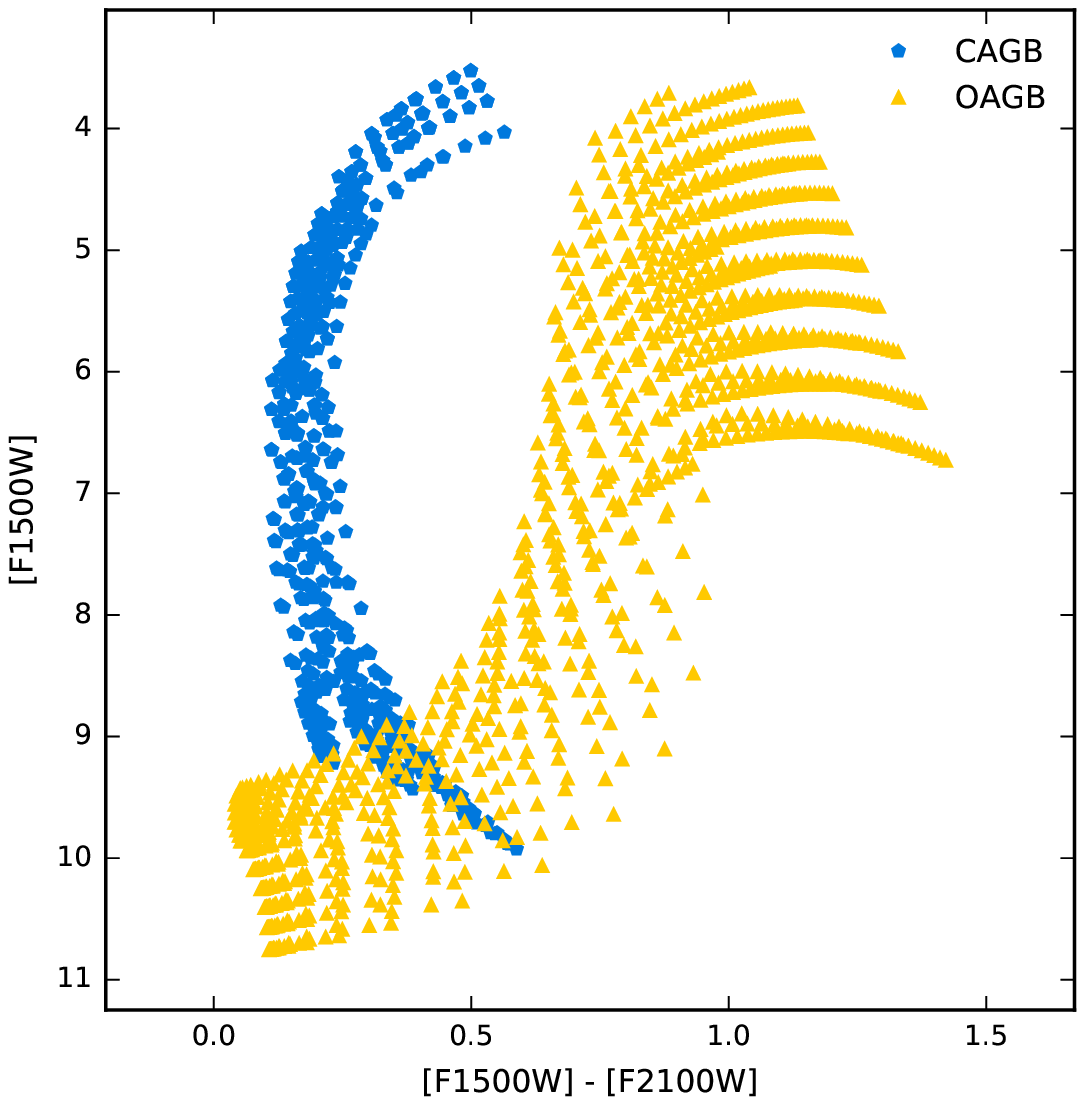}
\caption{Evolutionary tracks of AGB stars from the {\sc grams} models. Carbon-rich AGB stars (blue) occupy a narrow range in colour space. Oxygen-rich sources (yellow), have a wider spread in colour.  The points line up roughly in rows and/or columns, corresponding to sets of models with the same T$_{\rm eff}$, $R_{\rm in}$, and luminosity but increasing mass-loss rate. As the mass-loss increases, the AGB stars appear brighter and then redder in the CMDs. Only models with a stellar effective temperature of  T$_{\rm eff}$ = 2100--4700 K, a dust shell inner radius of R$_{\rm in}$ = 3 or 7 R$_{\rm star}$ and log(g) = -0.5 are shown. The models are set at a distance of 50 kpc. These CMDs can be directly compared to the observed sample in Figs.~\ref{fig:MIRICMD}B and \ref{fig:MIRICMD}F.}
\label{fig:MIRICMD_GRAMS}
\end{figure*}

\subsubsection{Oxygen-rich AGB Stars}

Stars on the early AGB are characterized by a slight IR excess, molecular absorption features and photometric variability. They do not show any prominent dust features and have MIRI colours slightly to the red of main-sequence stars. Around oxygen-rich stars, the fundamental vibrational mode of SiO causes a slight absorption feature in the spectrum at $\sim$8 $\mu$m, which will affect the [F770W] channel. 

 Simple metal oxides are the first dust species to form in low-density winds, with amorphous silicates becoming increasing prominent as the star evolves.  In O-rich stars with weak dust emission, a broad, low-contrast alumina (Al$_2$O$_3$) feature which peaks at $\sim$11 microns is often found to co-exist with amorphous silicates.  Therefore it maybe possible to trace changes in O-rich dust evolution using the four MIRI bands that cover the 10 $\mu$m region.

Amorphous silicates are the dominant dust component in oxygen-rich AGB stars, RSGs and post-AGB stars, with broad features at 10 and 20~$\mu$m. These features show considerable variation in shape, width and in the peak wavelength as the star evolves \citep[e.g.][]{SloanPrice1998, Cami1998, Speck2000, Jones2014}. This can cause a spread in the MIRI colours.  Oxygen-rich AGB stars are best identified using the [F1000W], [F1280W], [F1500W] and [F2100W] filters, e.g.~Fig.~\ref{fig:MIRICMD}, panels B, E and F.

 Fig.~\ref{fig:MIRICMD_GRAMS} shows the evolutionary tracks from the oxygen-rich  and carbon-rich {\sc grams} models. At low dust mass-loss rates, the O-rich {\sc grams} models form a narrow sequence in the [F1000W] vs.~[F1000W]--[F1500W] CMD, this reaches the highest magnitudes at the bluest colours. As the infrared excess increases the colours of the O-AGB stars start to diverge. They become redder and have fainter magnitudes. A similar dispersion in the mid-IR colour is not see in the carbon-stars with the onset of mass-loss; instead C-AGB stars tend to occupy a distinctive wedge  of less than 0.5 mag in IR colour space. 
 
The colours and evolutionary tracks of the AGB stars from the {\sc grams} models shows good agreement with the {\em JWST}/MIRI colours derived from IRS observation (c.f.~Fig.~\ref{fig:MIRICMD}B).  The major difference between the two results is the density of O-AGB sources with [F1000W]--[F1500W] $>$ 1; this is a consequence of the sampling in the models. The {\sc grams} models have not been weighted by an initial mass function, thus this region of colour space is over populated for a given stellar population. 
 
 O-AGB stars and C-AGB stars show the cleanest separation from each other in the [F1500W]--[F2100W] colour. In this diagram the {\sc grams} O-AGB stars have a higher source density at [F1500W]--[F2100W] $\sim$ 1.2 where the evolutionary tracks turn back on themselves; this is due to the source becoming optically thick and hence the silicate emission features changing to absorption features.

In some cases if the O-AGB star is undergoing an intense superwind and the dust shell is optically thick, the 10 $\mu$m silicate feature may be in absorption or self-absorption. These objects are rare, especially in low metallicity environments  \citep{Sloan2008, Jones2012, Jones2014} and thus  may blend in with other objects e.g.~YSOs in colour space.  For certain colour combinations, such as the [F1500W]--[F2100W] colour, an O-AGB star may have progressively redder colours until the source becomes optically thick. At this point, the colour now becomes bluer as mass-loss increases, and the star traverse back on itself in colour space. This effect is seen in the left panel of Fig.~\ref{fig:MIRICMD_GRAMS}.

\subsubsection{Oxygen-rich post-AGB stars}

Depending on the binary fraction of the star or the inclination angle of the disk, the SED of post-AGB stars can be either single peaked or double-peaked \citep[e.g.~][]{Ueta2003, vanAarle2011}. In post-AGB stars with a double-peaked SED; one peak is due to stellar emission from the hot central star and the other due to an outward moving circumstellar dust shell \citep{VanWinckel2003}. As post-AGB stars age, the detached, expanding dust shell cools and becomes fainter in the [F560W] and [F770W] bands, which are sensitive to warm dust \citep[e.g.][]{Min2013}. 

O-PAGB stars form a diagonal branch in the [F1000W]--[F1500W]~vs.~[F1000W] CMD  as seen in Fig.~\ref{fig:MIRICMD}B), where they occupy the less-dense region between C-rich AGB stars and YSOs. As they transition from the AGB region to the PN region of colour-space, O-rich PAGB stars can have similar mid-IR properties to evolved YSOs. To conclusively identify O-PAGB stars a long colour-baseline is required. The two colours should cover the hot stellar emission (with the NIRCam filters) and the second should be sensitive to the oxygen-rich dust.

Post-AGB stars with a binary companion can have a high crystalline silicate fraction and large dust grains indicative of a circumbinary dusty disc \citep{Gielen2011}. These post-AGB stars have a significant near-IR excess, which is superimposed on top of the stellar emission in the SED, rather than as double peak \citep{deRuyter2006}. In cases where the SED is singled peaked, O-rich post-AGB stars may be indistinguishable (in the MIRI filters) from their less evolved AGB counterparts. These `Disc' sources would be classified as AGB stars with a large dust excess by \citet{Woods2011}.

\subsubsection{Red Supergiants}

Red supergiants (RSGs) have very similar dust characteristics to O-AGB stars and are almost indistinguishable in colour space.  Due to their higher mass (8--25 $M_{\odot}$) RSGs are in general more luminous than AGB stars, and may fall on a sequence that extends to bright magnitudes than the AGB population, however, care needs to be taken as this region is also inhabited by foreground stars. 

Previous colour classification schemes for separating RSG from O-AGB stars rely on near-IR photometry \citep[e.g.][]{Boyer2011}. RSG (with little to no dust) are slightly bluer than the O-AGB stars in the $J-K_s$ versus $K_s$ and $J-$[3.6] CMDs due to their warmer effective temperatures. As RSG become enshrouded by dust, their $J-K_s$ colour grows redder and it becomes impossible to separate dusty AGB from RSG using the current set of photmetric bands.

Figure~\ref{fig:MIRICMD}  panel B shows that in the mid-IR RSGs are discernible by their [F1000W]-[F1500W] colours. In this CMD O-rich AGB stars form a left-leaning vertical sequence, which is slightly to the blue of a near-vertical sequence at [F1000W]-[F1500W] $\sim$ 0.2 which extends to bright magnitudes, formed of RSGs. Finally slightly to the red of this RSG track is a right-leaning vertical-sequence composed of C-AGB stars; all three sequences merge at lower luminosities where the photosphere dominate the colours in systems.  


\subsubsection{Carbon-Rich Stars}

In carbon-rich AGB stars, molecular absorption bands due to acetylene (C$_2$H$_2$) and numerous other carbon chain molecules form. Carbon rich molecules produce strong absorption features in the 4--8.5 and 13--14 $\mu$m wavelength intervals \citep{Matsuura2006}, the strength of these bands increases at low-metallicity which may affect the [F560W], [F770W] and [F1280W]  MIRI bands in metal poor carbon stars.  At solar metallicity, HCN may also be an important opacity source in carbon stars up to T$_{\rm eff}$ $\sim$ 2800 K \citep{Eriksson1984, Aoki1999, Harris2002}, affecting the [F560W] and [F1500W] MIRI fluxes.

Amorphous carbon and graphite are the dominant dust species in carbon stars  \citep[see][and references therein]{Groenewegen2009, Sloan2016}, they produce a dust-dominated continuum in the mid-IR, however they do not have a clear spectroscopic signature. Instead dust features due to Silicon Carbide (SiC) at 11.3 $\mu$m and the broad 26--30 $\mu$m feature,  which is often attributed to MgS dust  \citep{Goebel1985, Hony2002}, are the characteristic features of carbon-rich sources.

C-AGB stars are easily separated from all other types of objects using MIRI fluxes. The [F1500W]--[F2100W] colour (Fig.~\ref{fig:MIRICMD}F) is especially good for this as it separates carbon- and oxygen-rich evolved stars, and evolved stars from YSOs. Colours in the $\lambda = 10-20$ $\mu$m region enable us to separate these evolved stars as both carbon-rich and oxygen-rich AGB stars have prominent dust features in this region due to Silicon Carbide (SiC) and amorphous silicates, respectively.

\subsubsection{Carbon-rich post-AGB stars}

Carbon-rich post-AGB stars have UV and  optically excited PAH features, triangular SiC features at 11.3 $\mu$m, a prominent feature at 30 $\mu$m possibly due to MgS \citep{Sloan2007, Smolders2010, Matsuura2014}, although alternate carriers have been suggested \citep{Zhang2009} and in some instances a `unidentified' 21-$\mu$m emission feature which is unique to this class of stars \citep{Kwok1989, Hrivnak2009}. The strongest observed PAH features occur at wavelengths of 5.7, 6.2, 7.7, 8.6, and 11.3 $\mu$m.

C-PAGB stars occupy the relatively isolated region of colour-space between AGB stars and YSOs, thus they can be identified in MIRI colour-colour space using a wide variety of colours (see Figure~\ref{fig:MIRICCD},  panel C). By using a combination of filters pollution of the colour-selected C-PAGB stars by YSOs and extremely-red carbon stars is limited. 


\subsubsection{Planetary nebulae}

Planetary nebulae are emission line sources that can have a dust continuum that peaks in the mid-IR between $\lambda \sim 20-40$ $\mu$m, they may also have dust features due to either silicate or carbonaceous material (i.e.~SiC or MgS).  All the PNe in our sample have a thermal IR continuum which rises toward longer wavelengths, however the strength of the dust continuum compared to the emission lines varies considerably \citep{Stanghellini2007}. In addition to forbidden line emission from collisionally excited atomic species, most carbon-rich PNe (C-PN) have PAH emission lines in their spectra, with features at 6.2, 7.7 and 11.2 \mum typical \citep{BernardSalas2009}. PAH emission is often used as a discriminant between C-PN and O-PN as not all objects show solid state features.

Oxygen-rich PNe have spectral energy distributions that rapidly rise toward the far-IR. There is some overlap in colour with massive YSOs and background galaxies, however the O-PNe tend to be fainter than the vast majority of YSOs. Both the [F1130W]--[F1800W] and the [F1130W]--[F1500W] colours (see Figures~\ref{fig:MIRICMD}D and \ref{fig:MIRICMD}G) provide a clean separation between O-PN, YSOs and other emission line sources. The CMDs for both colours are almost identical, however the [F1130W]--[F1800W] colour has a slightly larger spread due to the longer baseline, and is thus preferred.

Due to their atomic emission lines, PAH features and rising dust continuum, carbon-rich PNe have colours which are more or less indistinguishable from the bulk of the YSO population. It is difficult to separate PNe and YSOs using only one or two colour combinations, however multiple CCD composed of four different MIRI bands may help in their identification (c.f.~Fig.~\ref{fig:MIRICCD}D). 


\subsubsection{Young Stellar Objects: YSOs}

As YSO's evolve their envelopes become hotter and less dense. Thus, YSOs of different evolutionary stages span a range in IR colours.  We illustrate this by comparing two independent but complementary YSO classification schemes; one from observed spectra \citep{Woods2011} and one from models \citep{Robitaille2006}. The spectral data provide the most accurate information for massive (M $>$ 8 M$_{\odot}$) YSOs, whilst the \citet{Robitaille2006} models provide the complete range of YSO masses down to 0.2 M$_{\odot}$.  Both these schemes portray a evolutionary sequence (1--3) from least to most evolved.

In the mid-IR, YSOs are characterized by a superposition of ice, PAH and oxygen-rich dust features on a very red cold dust continuum.  As discussed in Section~\ref{sec:AGB_YSO_obs},  we have divided the YSO spectra into four groups to reflect changes in their IR properties as they evolve towards the main sequence. The YSO spectral groups 1--3 represent an evolutionary sequence for massive YSOs: from deeply embedded sources in the early stages of formation, to stars surrounded by a dusty envelope which is being progressively dispersed. The majority of these sources would be classed as Stage I or Stage II YSOs on the colour-colour diagram of \citet{Robitaille2006}.

YSOs are prominent in all the MIRI bands. They occupy the reddest regions of colour space and are clearly separated from the evolved stars. Embedded protostellar objects (YSO-1) have the reddest colours, due to their dense cool envelopes. In principal as you go to bluer colours the more evolved a YSO will be. However, clear subdivisions between the YSO 1-3 categories is not possible, as YSOs inhabit regions with complex backgrounds. 
Evolved intermediate-mass YSOs (YSO-4) have a flat or declining spectrum with silicate emission, they occupy a slightly different region of colour space compared to the YSO 1--3 classes. This region of colour space is relatively sparse, typically populated by sources with dusty disks. 

The best way to isolate YSOs in colour space is to select two MIRI filters (e.g.~[F770W]--[F2500W] or [F1000W]--[F2100W]; Fig.~\ref{fig:MIRICMD}, panels A and C and Fig.~\ref{fig:MIRICCD}, panel C) with a wide baseline in wavelength that have $\lambda > 12$ $\mu$m. This mitigates contamination from AGB stars, which unlike YSOs only have moderate amounts of cold dust \citep{Jones2015b}.  Consequently, the YSOs have a rising SED at $\lambda > 20$ $\mu$m, whereas AGB stars and post-AGB stars generally have a falling spectra after 20$\mu$m.

\begin{figure*}
\centering
\includegraphics[width=80mm]{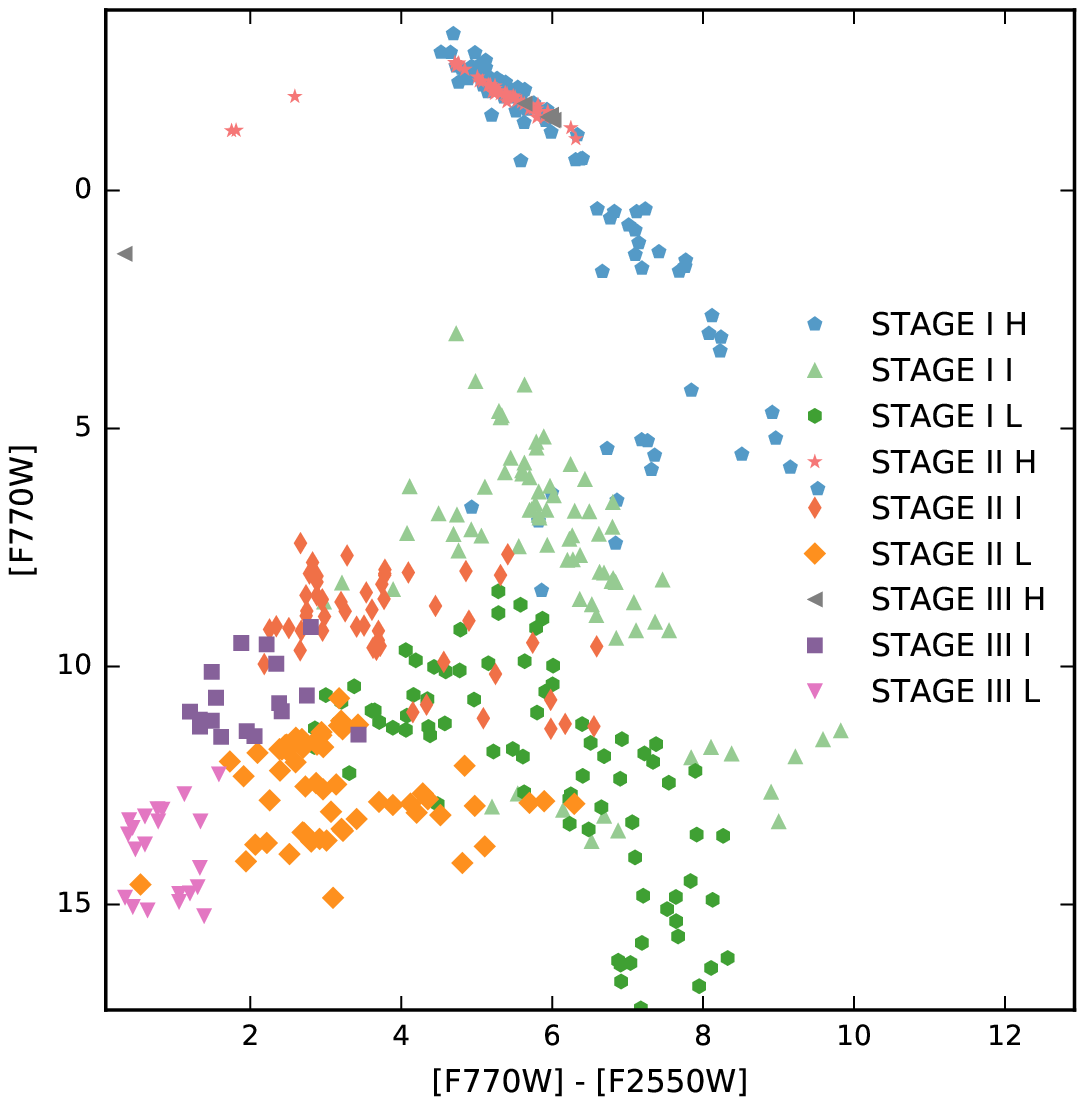}
\includegraphics[width=80mm]{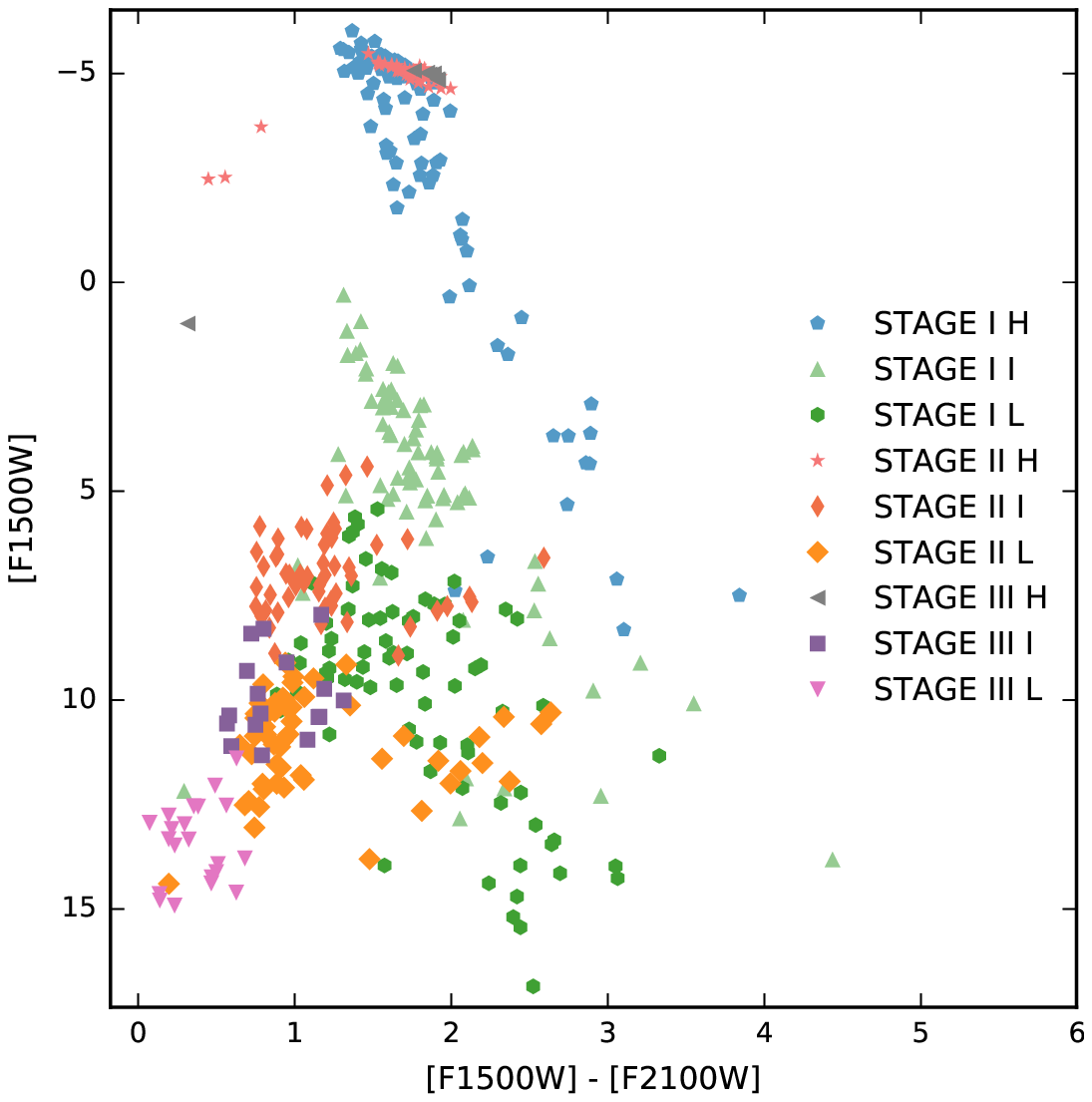}
\caption{ Selected CMDs from the YSO models and classification scheme by \citet{Robitaille2006} separated according to their mass and stage of evolution. The models are at a distance of  50 kpc. The L, I and H designations in the legend represent stars of 0.2, 2.0 and 20 $M_{\odot }$, respectively. For the low and intermediate mass YSOs a divisions between sources at different evolutionary stages is seen. The massive YSOs (20 $M_{\odot }$) have less definition. The distribution in luminosity is due to the initial stellar mass, with the most massive sources occupying the brighter regions of the CCDs. }
\label{fig:YSOCMD}
\end{figure*}

 Shown in Figure~\ref{fig:YSOCMD} are model sequences for YSOs with a stellar mass of 0.2, 2.0 and 20 $\pm 2.5\%$ $M_{\odot }$ at a disk inclination of 48.5$^{\circ}$. These models sample a variety of the SEDs in the \citet{Robitaille2006} model grid for low, intermediate and high mass stars. We identify in the MIRI CMDs and CCDs the models which correspond to each of the three evolutionary stages: Stage I, II and III defined by \citet{Robitaille2006}.  At earlier stages the disk geometry and the inclination has a significant effect on the SED, since this alters the optical depth along the line of sight.  Although, globally the inclination angle is not as important as it averages out over the stellar population. 
  Stage I sources have the reddest colours and correspond to the youngest embedded YSOs; here the envelope dominates the mid-IR flux.  As a source of a given mass evolves the envelope and disk disperse; the YSO becomes brighter with bluer colours due to the lower-extinction. An evolutionary sequence is seen in the colour distribution with  Stage III, II, and I sources moving from lower left to upper right.  Figure~\ref{fig:YSOCMD} shows that the evolutionary stages of YSOs are best separated with a long baseline with one colour $\gtrsim 18$  $\mu$m. This long baseline also provides the greatest separation between YSOs and non-YSOs.
  
It can be clearly seen in Figure~\ref{fig:YSOCMD} that the YSOs models have a large spread in luminosity, this corresponds to the initial mass of the YSO. In principal if you have a nearby field-of-view with little contamination you could cleanly separate the stages, and potentially detect YSOs with sub-solar masses. This would require substantial integrations times, even for the Magellanic Clouds.

\subsubsection{\Hii regions}

Compact \Hii regions form around young hot stars and are typically embedded in giant molecular clouds. They have a steeply rising infrared continuum and can be differentiated from PNe using the long-wavelength shape of the SED. Furthermore, \Hii regions are typically brighter than PNe and occupy the upper-right region of the MIRI CMDs. \Hii regions are among the reddest sources in the [F1000W]--[F1500W] CMD  (Fig.~\ref{fig:MIRICMD}B) due to their sharply rising continuum.

The [F1280W]--[F1800W] colour (Figs.~\ref{fig:MIRICMD} and \ref{fig:MIRICCD}, panels E and B, respectively) is one of the best filter selections for identifying \Hii regions. The limited overlap with a small number of YSOs in this colour space is probably due to the limited angular resolutions of {\em Spitzer's} IRS  which caused ambiguity in the spectral classification between (ultra)compact \Hii regions and evolved YSOs; consequently (ultra)compact \Hii regions were classified as YSOs by \cite{Woods2011}.


\subsubsection{Background Galaxies}

Background Galaxies are major contaminants to studies of IR stellar populations in the Local Group, particularly at low flux levels.  In the {\em JWST} era this problem will be further exacerbated as we push resolved stellar population studied to greater distances. One way of excluding galaxies to examine each individual source in the image to see if it is marginally resolved,  this can be done by in a quantitative manner by fitting an elliptical shape to the source. 

The constituents and morphologies of the individual galaxies in the background sample are described by \citet{Brown2014}. These galaxies cover a wide ranges in classes and have spectra which are composed of the composite stellar population for that galaxy. 
Quiescent galaxies tend to have blue colours and occupy a similar region to dust-free stars in MIRI CMDs.  Conversely, galaxies undergoing active star-formation populate a large range of MIRI colour space; due to their strong 6.2 and 7.7 $\mu$m PAH features and their steeply rising SEDs. As such star-forming galaxies have overlapping colours with YSOs. 

The MIRI CMDs and CCDs highlights the difficulty in distinguishing between YSOs and unresolved background sources in the same color-magnitude space; nonetheless, a careful choice of filter combinations can be used to separate galaxies from other stellar populations. Many classes of extragalatic objects will be quite red in the MIRI bands, and we find that the [F770W]--[F1000W] versus [F1500W]--[F2550W] CCD (Fig.~\ref{fig:MIRICCD}, panel D) is good discriminant between galactic and non-galactic sources. This CCD also highlights the diversity in galaxy colours. 



\section{Applicability to Local Volume galaxies} \label{sec:LocalVol}

{\em JWST}/MIRI's unparalleled sensitivity and spatial resolution at 5--28 $\mu$m will enable resolved stellar population studies out to $\sim$4 Mpc; over 100 galaxies fall within this volume. Going beyond the Local Group with detailed studies of resolved stellar populations will enable us to measure star formation histories, probe active star formation regions, ascertain the age of the oldest stellar populations and determine the chemical and dust enrichment of galaxies with properties very different to our own. 

Local volume galaxies span a wide range in metallicity ($-2.72$  $<$  [Fe/H] $<$  0.5). This large metallicity baseline provides a foundation for understanding which types of objects produce dust, and the significance of their dust production as galaxies evolve. Insights into the connection between resolved stellar populations and galaxy evolution can be used to study the early universe. For instance, it is unclear if intermediate- and high-mass stars can account for the substantial dust abundances observed at $z \gtrsim 6.4$ \citep{Beelen2006,  Valiante2009, Gall2011}.


\begin{figure}
\centering
\includegraphics[width=84mm]{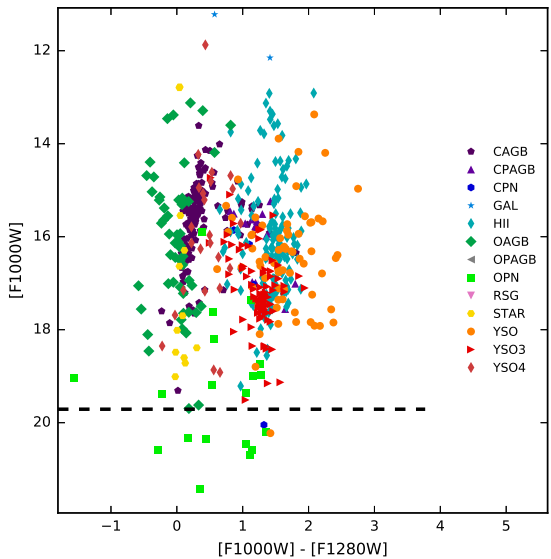}
\caption{The LMC placed at a distance of 3.6 Mpc appropriate for M81. A 10$\sigma$ detection in 10,000s at [F1000W] is marked as a dashed line.}
\label{fig:MIRICMD_scaled}
\end{figure}

\begin{table}
\begin{center}
\caption{The 10$\sigma$ 10000s magnitude limits for the MIRI filters, estimated from the zero points given in Table~\ref{tab:Zpoints}.}
\label{tab:MagLimits}
\begin{tabular}{lc}
\hline
\hline
Filter  &   Magnitude     \\
\hline
F560W	&	22.15	\\	
F770W	&	21.09	\\	
F1000W	&	19.69	\\	
F1130W	&	18.43	\\	
F1280W	&	18.67	\\	
F1500W	&	17.78	\\	
F1800W	&	16.39	\\	
F2100W	&	15.27	\\	
F2550W	&	13.43	\\	
\hline
\end{tabular}
\end{center}
\end{table}


{\em JWST} will be able to obtain moderately-deep multi-band stellar photometry for the M81 group at 3.6 Mpc and the Sculptor filament at 3.9 Mpc.   In Table~\ref{tab:MagLimits} we list the 10$\sigma$ magnitude limits for long-duration observations of faint sources with the MIRI filters.
In Figure~\ref{fig:MIRICMD_scaled} we have scaled the MIRI photometry of the LMC to that of a galaxy at 3.6 Mpc and over-plotted the MIRI sensitivity limits for a 10$\sigma$ detection in 10,000s on the [F1000W] vs.~[F1000W]--[F1280W] CMD. At this distance only the brighter red objects, such as RSGs, AGB stars, PNe, massive YSOs (M $\gtrsim$ 10 M$_{\odot}$) and \Hii regions can be detected.  The depth/detection limit of {\em JWST} observations of galaxies at 3 Mpc would be comparable to those of the Wide-Field Infrared Survey Explorer (WISE) for the LMC \citep[e.g.][]{Wright2010, Nikutta2014}. From these observations we would be able to study the formation of new stars, characterize the largest dust producers and probe substructures within a galaxy.



IR surveys similar to the {\em Spitzer} SAGE \citep{Meixner2006, Gordon2011} legacy programs which conducted census of all objects in the Magellanic Clouds brighter than 15 mag at 8 $\mu$m are achievable (in 10,000s per filter, per pointing) with {\em JWST} out to $\sim$450 kpc. 
The decline in source density towards fainter magnitudes in  Figure~\ref{fig:MIRICMD_scaled} is a limitation of the {\em Spitzer} spectroscopic sample, rather than an observation effect which would be due to {\em JWST}/MIRI. Colours of fainter populations e.g. low-mass YSOs and evolved stars can be inferred from scaling the models presented in Section~\ref{sec:results} to the required distance and comparing these to the sensitivity limits derived in Table~\ref{tab:MagLimits}.

 MIRI observations would allow detailed studies of individual YSOs and H\,{\sc ii} regions. YSOs can be identified with only MIRI two colours; however multi-band photometry  is required to constrain the physical processes in the protostar, circumstellar disk, and collapsing outer envelope via radiative transfer modelling of its SED.
By observing red-luminous populations we can also determine star formation rates for a much larger volume ($\gtrsim$100) of galaxies than is currently possible with {\em Spitzer} \citep[e.g][]{Whitney2008, Sewilo2013}. These studies extrapolate the mass function derived from star counts and SED fitting with a standard IMF, to derive an empirical limit for the current star formation rate of a galaxy. Furthermore, MIRI observation of YSOs in conjunction with H\,{\sc i}, H$\alpha$, and CO gas tracers will reveal the the initial conditions and process of star-formation across a range of galactic environments.

The IR emission from evolved stellar populations of local volume galaxies measured by {\em JWST} can be compared to the ISM dust masses obtained from global flux measurements  \citep{Draine2007} by the {\em Spitzer} Infrared Nearby Galaxies Survey \citep[SINGS;][]{Kennicutt2003}.  Determining the nature of interstellar dust as galaxies evolve in metallicity is essential, as its origin and rate of destruction by supernova shocks \citep[e.g][]{Temim2015} is uncertain. Dust production rates of evolved stars can be estimated from mid-IR colours \citep{Matsuura2009, Jones2015a} or via SED fitting with model grids \citep{Riebel2012, Srinivasan2016}. This can constrain the luminosities, dust chemistry and the current dust-production rate of each star in the sample.  The rate of dust injection by the evolved stellar population can be compared to the current ISM dust mass to determine the time-scale at which stellar sources replenish ISM dust.
This will directly constrain dust models \citep[e.g.][]{Dwek2011} and provide the foundations for deciphering {\em JWST} observations of more distant galaxies in the high-redshift universe.

To ensure representation, samples selected based on photometric data needs to be confirmed with spectroscopic observations. 
With {\em JWST} we will be able to conduct detailed spectroscopic studies of point sources within the Local Group. Spectroscopic surveys will provide critical information about the dust compositions,  spectral types and ages of each star. 
By linking the MIRI colours of LMC objects to their dust characteristics and infrared spectral type (see Section~\ref{sec:MIRIcolours}) we can effectively distinguish between sources of similar colour and efficiently select candidates for {\em JWST} spectroscopic observations through the careful use of the MIRI CMDs. This will ensure that there is a high success rate of {\em JWST} spectroscopically observing targets with the desired object class and chemistry.  Furthermore, spectroscopic observations with {\em JWST} can evaluate and refine our proposed photometric classification scheme, to ensure clean subdivisions between stellar classes.

\section{Summary} \label{sec:conclusion}

{\em JWST} is expected to revolutionize our understanding of composite stellar populations in the Local Volume. 
In this paper we have used over 1250 {\em Spitzer}-IRS spectra from the SAGE-Spec legacy program and model spectra from the {\sc grams} and YSO grids to calculate synthetic MIRI fluxes and magnitudes for IR-bright stellar populations and background galaxies. 
These results can be used to select {\em JWST} filters which will provide good photometric sampling of SEDs, used to measure the stellar and circumstellar properties of the stars. 

We have developed colour-colour and colour-magnitude classification schemes for {\em JWST}/MIRI to identify and select samples of similar objects for future studies. These results can easily be adapted for use in other galaxies, e.g.~the M81 group, to probe composite stellar populations across a range of galaxy types, metallicities and star formation environments, beyond the boundaries of the Local Group. 
In Section~\ref{sec:MIRIcolours} we assessed the best filters to target specific stellar populations throughout a galaxy, and discussed how to remove foreground and background contamination based on their {\em JWST} broadband colours. We highlight the F1000W and F2100W MIRI filters as they provide a clean separation between evolved stars, young stellar objects and background galaxies.

 \vspace{0.4cm}

Jones and Meixner acknowledge support from NASA grant, NNX14AN06G, for this work. Justtanont is partly supported by the Swedish National Space Board.

\appendix

\section{Flux calibration of the {\em Spitzer} spectra}


 Prior to joining the different modules of the {\em Spitzer} IRS spectra together, flux uncertainties in the {\em Spitzer} sample are carefully propagated using formal uncertainties in the mean ($\sigma /\sqrt{N}$). To photometrically calibrate the spectra, the mean flux density is converted to Jy using IRS observations of the standard stars HR 6348 (K0 III) for the short-low data, and HR 6348 and HD 173511 (K5 III) for the long-low \citep{Sloan2015a}. Spectra observed with the short-high and long-high modules were extracted using a full-slit extraction and calibrated using $\xi$ Dra (K2 III) as a standard. 

 The flux calibration was applied to each individual spectral nod and order. Here the spectroscopic (i.e.~the point-to-point) uncertainty in the flux is better than 0.5\% at most wavelengths \citep{Sloan2015a}. When combining the spectra from the nod positions, the larger of the propagated uncertainties from the two nods or the uncertainty in mean was adopted. This combination introduces several systematic errors in the flux \citep{Lebouteiller2011}, which depends on source geometry (G.~Sloan 2016, private communication). The error in the absolute flux calibration is generally around $\sim$5\%, however this can be considerably more for extended sources or bad pointings.

To produce the final spectrum, a scalar multiplicative correction was applied to each spectral segment to remove discontinuities which arise from pointing errors. Segments are normalised upwards (using the wavelengths where they overlapped), to align with the best-centered segment. This correction is typically less than 10 \% and has no dependence on wavelength. Data at the ends of the segment that could not be calibrated reliably were trimmed from the spectra. As with the low-resolution modules, short-high data was stitched to the long-high data, but no orders were adjusted relative to other orders within the same module, as they were obtained at the same time with identical telescope pointings. For more details on this procedure see \citet{Woods2011b, Lebouteiller2011, Sloan2015a}. \\







\def\aj{AJ}					
\def\actaa{Acta Astron.}                        
\def\araa{ARA\&A}				
\def\apj{ApJ}					
\def\apjl{ApJL}					
\def\apjs{ApJS}					
\def\ao{Appl.~Opt.}				
\def\apss{Ap\&SS}				
\def\aap{A\&A}					
\def\aapr{A\&A~Rev.}				
\def\aaps{A\&AS}				
\def\azh{AZh}					
\def\baas{BAAS}					
\def\jrasc{JRASC}				
\def\memras{MmRAS}				
\def\mnras{MNRAS}				
\def\pra{Phys.~Rev.~A}				
\def\prb{Phys.~Rev.~B}				
\def\prc{Phys.~Rev.~C}				
\def\prd{Phys.~Rev.~D}				
\def\pre{Phys.~Rev.~E}				
\def\prl{Phys.~Rev.~Lett.}			
\def\pasp{PASP}					
\def\pasj{PASJ}					
\def\qjras{QJRAS}				
\def\skytel{S\&T}				
\def\solphys{Sol.~Phys.}			
\def\sovast{Soviet~Ast.}			
\def\ssr{Space~Sci.~Rev.}			
\def\zap{ZAp}					
\def\nat{Nature}				
\def\iaucirc{IAU~Circ.}				
\def\aplett{Astrophys.~Lett.}			
\def\apspr{Astrophys.~Space~Phys.~Res.}		
\def\bain{Bull.~Astron.~Inst.~Netherlands}	
\def\fcp{Fund.~Cosmic~Phys.}			
\def\gca{Geochim.~Cosmochim.~Acta}		
\def\grl{Geophys.~Res.~Lett.}			
\def\jcp{J.~Chem.~Phys.}			
\def\jgr{J.~Geophys.~Res.}			
\def\jqsrt{J.~Quant.~Spec.~Radiat.~Transf.}	
\def\memsai{Mem.~Soc.~Astron.~Italiana}		
\def\nphysa{Nucl.~Phys.~A}			
\def\physrep{Phys.~Rep.}			
\def\physscr{Phys.~Scr}				
\def\planss{Planet.~Space~Sci.}			
\def\procspie{Proc.~SPIE}			
\def\icarus{Icarus}
\let\astap=\aap
\let\apjlett=\apjl
\let\apjsupp=\apjs
\let\applopt=\ao


\bibliographystyle{aa}


\begin{thebibliography}{97}
\expandafter\ifx\csname natexlab\endcsname\relax\def\natexlab#1{#1}\fi

\bibitem[{Adams {et~al.}(2013)Adams, Simon, Bolatto, Sloan, Sandstrom,
  Schmiedeke, van Loon, Oliveira, \& Keller}]{Adams2013}
Adams, J.~J., Simon, J.~D., Bolatto, A.~D., {et~al.} 2013, The Astrophysical
  Journal, 771, 112

\bibitem[{{Aoki} {et~al.}(1999){Aoki}, {Tsuji}, \& {Ohnaka}}]{Aoki1999}
{Aoki}, W., {Tsuji}, T., \& {Ohnaka}, K. 1999, \aap, 350, 945

\bibitem[{{Aringer} {et~al.}(2009){Aringer}, {Girardi}, {Nowotny}, {Marigo}, \&
  {Lederer}}]{Aringer2009}
{Aringer}, B., {Girardi}, L., {Nowotny}, W., {Marigo}, P., \& {Lederer}, M.~T.
  2009, \aap, 503, 913

\bibitem[{{Beelen} {et~al.}(2006){Beelen}, {Cox}, {Benford}, {Dowell},
  {Kov{\'a}cs}, {Bertoldi}, {Omont}, \& {Carilli}}]{Beelen2006}
{Beelen}, A., {Cox}, P., {Benford}, D.~J., {et~al.} 2006, \apj, 642, 694

\bibitem[{{Beichman} {et~al.}(1988){Beichman}, {Neugebauer}, {Habing}, {Clegg},
  \& {Chester}}]{Beichman1988}
{Beichman}, C.~A., {Neugebauer}, G., {Habing}, H.~J., {Clegg}, P.~E., \&
  {Chester}, T.~J., eds. 1988, {Infrared astronomical satellite (IRAS) catalogs
  and atlases. Volume 1: Explanatory supplement}, Vol.~1

\bibitem[{{Bernard-Salas} {et~al.}(2009){Bernard-Salas}, {Peeters}, {Sloan},
  {Gutenkunst}, {Matsuura}, {Tielens}, {Zijlstra}, \&
  {Houck}}]{BernardSalas2009}
{Bernard-Salas}, J., {Peeters}, E., {Sloan}, G.~C., {et~al.} 2009, \apj, 699,
  1541

\bibitem[{{Bernard-Salas} {et~al.}(2008){Bernard-Salas}, {Pottasch},
  {Gutenkunst}, {Morris}, \& {Houck}}]{BernardSalas2008}
{Bernard-Salas}, J., {Pottasch}, S.~R., {Gutenkunst}, S., {Morris}, P.~W., \&
  {Houck}, J.~R. 2008, \apj, 672, 274

\bibitem[{{Bessell}(2000)}]{Bessell2000}
{Bessell}, M.~S. 2000, \pasp, 112, 961

\bibitem[{{Blommaert} {et~al.}(2003){Blommaert}, {Siebenmorgen}, {Coulais},
  {Metcalfe}, {Miville-Desch{\^e}nes}, {Okumura}, {Ott}, {Pollack}, {Sauvage},
  \& {Starck}}]{Blommaert2003}
{Blommaert}, J., {Siebenmorgen}, R., {Coulais}, A., {et~al.}, eds. 2003, ESA
  Special Publication, Vol. 1262, {The ISO Handbook Volume II: CAM - The ISO
  Camera (v 2.0)}

\bibitem[{{Blum} {et~al.}(2006){Blum}, {Mould}, {Olsen}, {Frogel}, {Werner},
  {Meixner}, {Markwick-Kemper}, {Indebetouw}, {Whitney}, {Meade}, {Babler},
  {Churchwell}, {Gordon}, {Engelbracht}, {For}, {Misselt}, {Vijh}, {Leitherer},
  {Volk}, {Points}, {Reach}, {Hora}, {Bernard}, {Boulanger}, {Bracker},
  {Cohen}, {Fukui}, {Gallagher}, {Gorjian}, {Harris}, {Kelly}, {Kawamura},
  {Latter}, {Madden}, {Mizuno}, {Mizuno}, {Nota}, {Oey}, {Onishi}, {Paladini},
  {Panagia}, {Perez-Gonzalez}, {Shibai}, {Sato}, {Smith}, {Staveley-Smith},
  {Tielens}, {Ueta}, {Van Dyk}, \& {Zaritsky}}]{Blum2006}
{Blum}, R.~D., {Mould}, J.~R., {Olsen}, K.~A., {et~al.} 2006, \aj, 132, 2034

\bibitem[{{Bohlin} {et~al.}(2011){Bohlin}, {Gordon}, {Rieke}, {Ardila},
  {Carey}, {Deustua}, {Engelbracht}, {Ferguson}, {Flanagan}, {Kalirai},
  {Meixner}, {Noriega-Crespo}, {Su}, \& {Tremblay}}]{Bohlin2011}
{Bohlin}, R.~C., {Gordon}, K.~D., {Rieke}, G.~H., {et~al.} 2011, \aj, 141, 173

\bibitem[{{Bouchet} {et~al.}(2015){Bouchet}, {Garc{\'{\i}}a-Mar{\'{\i}}n},
  {Lagage}, {Amiaux}, {Augu{\'e}res}, {Bauwens}, {Blommaert}, {Chen}, {Detre},
  {Dicken}, {Dubreuil}, {Galdemard}, {Gastaud}, {Glasse}, {Gordon}, {Gougnaud},
  {Guillard}, {Justtanont}, {Krause}, {Leboeuf}, {Longval}, {Martin}, {Mazy},
  {Moreau}, {Olofsson}, {Ray}, {Rees}, {Renotte}, {Ressler}, {Ronayette},
  {Salasca}, {Scheithauer}, {Sykes}, {Thelen}, {Wells}, {Wright}, \&
  {Wright}}]{Bouchet2015}
{Bouchet}, P., {Garc{\'{\i}}a-Mar{\'{\i}}n}, M., {Lagage}, P.-O., {et~al.}
  2015, \pasp, 127, 612

\bibitem[{{Boyer} {et~al.}(2011){Boyer}, {Srinivasan}, {van Loon}, {McDonald},
  {Meixner}, {Zaritsky}, {Gordon}, {Kemper}, {Babler}, {Block}, {Bracker},
  {Engelbracht}, {Hora}, {Indebetouw}, {Meade}, {Misselt}, {Robitaille},
  {Sewi{\l}o}, {Shiao}, \& {Whitney}}]{Boyer2011}
{Boyer}, M.~L., {Srinivasan}, S., {van Loon}, J.~T., {et~al.} 2011, \aj, 142,
  103

\bibitem[{{Brown} {et~al.}(2014){Brown}, {Moustakas}, {Smith}, {da Cunha},
  {Jarrett}, {Imanishi}, {Armus}, {Brandl}, \& {Peek}}]{Brown2014}
{Brown}, M.~J.~I., {Moustakas}, J., {Smith}, J.-D.~T., {et~al.} 2014, \apjs,
  212, 18

\bibitem[{{Buchanan} {et~al.}(2006){Buchanan}, {Kastner}, {Forrest}, {Hrivnak},
  {Sahai}, {Egan}, {Frank}, \& {Barnbaum}}]{Buchanan2006}
{Buchanan}, C.~L., {Kastner}, J.~H., {Forrest}, W.~J., {et~al.} 2006, \aj, 132,
  1890

\bibitem[{{Buchanan} {et~al.}(2009){Buchanan}, {Kastner}, {Hrivnak}, \&
  {Sahai}}]{Buchanan2009}
{Buchanan}, C.~L., {Kastner}, J.~H., {Hrivnak}, B.~J., \& {Sahai}, R. 2009,
  \aj, 138, 1597

\bibitem[{{Cami} {et~al.}(1998){Cami}, {de Jong}, {Justtannont}, {Yamamura}, \&
  {Waters}}]{Cami1998}
{Cami}, J., {de Jong}, T., {Justtannont}, K., {Yamamura}, I., \& {Waters},
  L.~B.~F.~M. 1998, \apss, 255, 339

\bibitem[{{Cohen} {et~al.}(1992){Cohen}, {Walker}, {Barlow}, \&
  {Deacon}}]{Cohen1992}
{Cohen}, M., {Walker}, R.~G., {Barlow}, M.~J., \& {Deacon}, J.~R. 1992, \aj,
  104, 1650

\bibitem[{{de Ruyter} {et~al.}(2006){de Ruyter}, {van Winckel}, {Maas}, {Lloyd
  Evans}, {Waters}, \& {Dejonghe}}]{deRuyter2006}
{de Ruyter}, S., {van Winckel}, H., {Maas}, T., {et~al.} 2006, \aap, 448, 641

\bibitem[{{Draine} {et~al.}(2007){Draine}, {Dale}, {Bendo}, {Gordon}, {Smith},
  {Armus}, {Engelbracht}, {Helou}, {Kennicutt}, {Li}, {Roussel}, {Walter},
  {Calzetti}, {Moustakas}, {Murphy}, {Rieke}, {Bot}, {Hollenbach}, {Sheth}, \&
  {Teplitz}}]{Draine2007}
{Draine}, B.~T., {Dale}, D.~A., {Bendo}, G., {et~al.} 2007, \apj, 663, 866

\bibitem[{{Dwek} \& {Cherchneff}(2011)}]{Dwek2011}
{Dwek}, E. \& {Cherchneff}, I. 2011, \apj, 727, 63

\bibitem[{{Egan} {et~al.}(2001){Egan}, {Van Dyk}, \& {Price}}]{Egan2001}
{Egan}, M.~P., {Van Dyk}, S.~D., \& {Price}, S.~D. 2001, \aj, 122, 1844

\bibitem[{{Engelbracht} {et~al.}(2007){Engelbracht}, {Blaylock}, {Su}, {Rho},
  {Rieke}, {Muzerolle}, {Padgett}, {Hines}, {Gordon}, {Fadda},
  {Noriega-Crespo}, {Kelly}, {Latter}, {Hinz}, {Misselt}, {Morrison},
  {Stansberry}, {Shupe}, {Stolovy}, {Wheaton}, {Young}, {Neugebauer},
  {Wachter}, {P{\'e}rez-Gonz{\'a}lez}, {Frayer}, \&
  {Marleau}}]{Engelbracht2007}
{Engelbracht}, C.~W., {Blaylock}, M., {Su}, K.~Y.~L., {et~al.} 2007, \pasp,
  119, 994

\bibitem[{{Eriksson} {et~al.}(1984){Eriksson}, {Gustafsson}, {Jorgensen}, \&
  {Nordlund}}]{Eriksson1984}
{Eriksson}, K., {Gustafsson}, B., {Jorgensen}, U.~G., \& {Nordlund}, A. 1984,
  \aap, 132, 37

\bibitem[{{Feast}(2013)}]{Feast2013}
{Feast}, M.~W. 2013, {Galactic Distance Scales}, ed. T.~D. {Oswalt} \&
  G.~{Gilmore}, 829

\bibitem[{{Gall} {et~al.}(2011){Gall}, {Hjorth}, \& {Andersen}}]{Gall2011}
{Gall}, C., {Hjorth}, J., \& {Andersen}, A.~C. 2011, \aapr, 19, 43

\bibitem[{{Gielen} {et~al.}(2011){Gielen}, {Bouwman}, {van Winckel}, {Lloyd
  Evans}, {Woods}, {Kemper}, {Marengo}, {Meixner}, {Sloan}, \&
  {Tielens}}]{Gielen2011}
{Gielen}, C., {Bouwman}, J., {van Winckel}, H., {et~al.} 2011, \aap, 533, A99

\bibitem[{{Glasse} {et~al.}(2015){Glasse}, {Rieke}, {Bauwens},
  {Garc{\'{\i}}a-Mar{\'{\i}}n}, {Ressler}, {Rost}, {Tikkanen}, {Vandenbussche},
  \& {Wright}}]{Glasse2015}
{Glasse}, A., {Rieke}, G.~H., {Bauwens}, E., {et~al.} 2015, \pasp, 127, 686

\bibitem[{{Goebel} \& {Moseley}(1985)}]{Goebel1985}
{Goebel}, J.~H. \& {Moseley}, S.~H. 1985, \apjl, 290, L35

\bibitem[{{Gordon} {et~al.}(2011){Gordon}, {Meixner}, {Meade}, {Whitney},
  {Engelbracht}, {Bot}, {Boyer}, {Lawton}, {Sewi{\l}o}, {Babler}, {Bernard},
  {Bracker}, {Block}, {Blum}, {Bolatto}, {Bonanos}, {Harris}, {Hora},
  {Indebetouw}, {Misselt}, {Reach}, {Shiao}, {Tielens}, {Carlson},
  {Churchwell}, {Clayton}, {Chen}, {Cohen}, {Fukui}, {Gorjian}, {Hony},
  {Israel}, {Kawamura}, {Kemper}, {Leroy}, {Li}, {Madden}, {Marble},
  {McDonald}, {Mizuno}, {Mizuno}, {Muller}, {Oliveira}, {Olsen}, {Onishi},
  {Paladini}, {Paradis}, {Points}, {Robitaille}, {Rubin}, {Sandstrom}, {Sato},
  {Shibai}, {Simon}, {Smith}, {Srinivasan}, {Vijh}, {Van Dyk}, {van Loon}, \&
  {Zaritsky}}]{Gordon2011}
{Gordon}, K.~D., {Meixner}, M., {Meade}, M.~R., {et~al.} 2011, \aj, 142, 102

\bibitem[{{Groenewegen} {et~al.}(2009){Groenewegen}, {Sloan}, {Soszy{\'n}ski},
  \& {Petersen}}]{Groenewegen2009}
{Groenewegen}, M.~A.~T., {Sloan}, G.~C., {Soszy{\'n}ski}, I., \& {Petersen},
  E.~A. 2009, \aap, 506, 1277

\bibitem[{{Gruendl} \& {Chu}(2009)}]{Gruendl2009}
{Gruendl}, R.~A. \& {Chu}, Y.-H. 2009, \apjs, 184, 172

\bibitem[{{Harris} {et~al.}(2002){Harris}, {Polyansky}, \&
  {Tennyson}}]{Harris2002}
{Harris}, G.~J., {Polyansky}, O.~L., \& {Tennyson}, J. 2002, \apj, 578, 657

\bibitem[{{Hony} {et~al.}(2002){Hony}, {Waters}, \& {Tielens}}]{Hony2002}
{Hony}, S., {Waters}, L.~B.~F.~M., \& {Tielens}, A.~G.~G.~M. 2002, \aap, 390,
  533

\bibitem[{{Hora} {et~al.}(2008){Hora}, {Cohen}, {Ellis}, {Meixner}, {Blum},
  {Latter}, {Whitney}, {Meade}, {Babler}, {Indebetouw}, {Gordon},
  {Engelbracht}, {For}, {Block}, {Misselt}, {Vijh}, \& {Leitherer}}]{Hora2008}
{Hora}, J.~L., {Cohen}, M., {Ellis}, R.~G., {et~al.} 2008, \aj, 135, 726

\bibitem[{{Houck} {et~al.}(2004){Houck}, {Roellig}, {van Cleve}, {Forrest},
  {Herter}, {Lawrence}, {Matthews}, {Reitsema}, {Soifer}, {Watson}, {Weedman},
  {Huisjen}, {Troeltzsch}, {Barry}, {Bernard-Salas}, {Blacken}, {Brandl},
  {Charmandaris}, {Devost}, {Gull}, {Hall}, {Henderson}, {Higdon}, {Pirger},
  {Schoenwald}, {Sloan}, {Uchida}, {Appleton}, {Armus}, {Burgdorf},
  {Fajardo-Acosta}, {Grillmair}, {Ingalls}, {Morris}, \& {Teplitz}}]{Houck2004}
{Houck}, J.~R., {Roellig}, T.~L., {van Cleve}, J., {et~al.} 2004, \apjs, 154,
  18

\bibitem[{{Hrivnak} {et~al.}(2009){Hrivnak}, {Volk}, \& {Kwok}}]{Hrivnak2009}
{Hrivnak}, B.~J., {Volk}, K., \& {Kwok}, S. 2009, \apj, 694, 1147

\bibitem[{{Jones} {et~al.}(2012){Jones}, {Kemper}, {Sargent}, {McDonald},
  {Gielen}, {Woods}, {Sloan}, {Boyer}, {Zijlstra}, {Clayton}, {Kraemer},
  {Srinivasan}, \& {Ruffle}}]{Jones2012}
{Jones}, O.~C., {Kemper}, F., {Sargent}, B.~A., {et~al.} 2012, \mnras, 427,
  3209

\bibitem[{{Jones} {et~al.}(2014){Jones}, {Kemper}, {Srinivasan}, {McDonald},
  {Sloan}, \& {Zijlstra}}]{Jones2014}
{Jones}, O.~C., {Kemper}, F., {Srinivasan}, S., {et~al.} 2014, \mnras, 440, 631

\bibitem[{{Jones} {et~al.}(2015{\natexlab{a}}){Jones}, {McDonald}, {Rich},
  {Kemper}, {Boyer}, {Zijlstra}, \& {Bendo}}]{Jones2015a}
{Jones}, O.~C., {McDonald}, I., {Rich}, R.~M., {et~al.} 2015{\natexlab{a}},
  \mnras, 446, 1584

\bibitem[{{Jones} {et~al.}(2015{\natexlab{b}}){Jones}, {Meixner}, {Sargent},
  {Boyer}, {Sewi{\l}o}, {Hony}, \& {Roman-Duval}}]{Jones2015b}
{Jones}, O.~C., {Meixner}, M., {Sargent}, B.~A., {et~al.} 2015{\natexlab{b}},
  \apj, 811, 145

\bibitem[{{Kemper} {et~al.}(2010){Kemper}, {Woods}, {Antoniou}, {Bernard},
  {Blum}, {Boyer}, {Chan}, {Chen}, {Cohen}, {Dijkstra}, {Engelbracht},
  {Galametz}, {Galliano}, {Gielen}, {Gordon}, {Gorjian}, {Harris}, {Hony},
  {Hora}, {Indebetouw}, {Jones}, {Kawamura}, {Lagadec}, {Lawton}, {Leisenring},
  {Madden}, {Marengo}, {Matsuura}, {McDonald}, {McGuire}, {Meixner}, {Mulia},
  {O'Halloran}, {Oliveira}, {Paladini}, {Paradis}, {Reach}, {Rubin},
  {Sandstrom}, {Sargent}, {Sewilo}, {Shiao}, {Sloan}, {Speck}, {Srinivasan},
  {Szczerba}, {Tielens}, {van Aarle}, {Van Dyk}, {van Loon}, {Van Winckel},
  {Vijh}, {Volk}, {Whitney}, {Wilkins}, \& {Zijlstra}}]{Kemper2010}
{Kemper}, F., {Woods}, P.~M., {Antoniou}, V., {et~al.} 2010, \pasp, 122, 683

\bibitem[{{Kennicutt} {et~al.}(2003){Kennicutt}, {Armus}, {Bendo}, {Calzetti},
  {Dale}, {Draine}, {Engelbracht}, {Gordon}, {Grauer}, {Helou}, {Hollenbach},
  {Jarrett}, {Kewley}, {Leitherer}, {Li}, {Malhotra}, {Regan}, {Rieke},
  {Rieke}, {Roussel}, {Smith}, {Thornley}, \& {Walter}}]{Kennicutt2003}
{Kennicutt}, Jr., R.~C., {Armus}, L., {Bendo}, G., {et~al.} 2003, \pasp, 115,
  928

\bibitem[{{Kraemer} {et~al.}(2017){Kraemer}, {Sloan}, {Wood}, {Jones}, \&
  {Egan}}]{Kraemer2017}
{Kraemer}, K.~E., {Sloan}, G.~C., {Wood}, P.~R., {Jones}, O.~C., \& {Egan},
  M.~P. 2017, \apj, 834, 185

\bibitem[{{Ku{\v c}inskas} {et~al.}(2005){Ku{\v c}inskas}, {Hauschildt},
  {Ludwig}, {Brott}, {Vansevi{\v c}ius}, {Lindegren}, {Tanab{\'e}}, \&
  {Allard}}]{Kucinskas2005}
{Ku{\v c}inskas}, A., {Hauschildt}, P.~H., {Ludwig}, H.-G., {et~al.} 2005,
  \aap, 442, 281

\bibitem[{{Kwok} {et~al.}(1989){Kwok}, {Volk}, \& {Hrivnak}}]{Kwok1989}
{Kwok}, S., {Volk}, K.~M., \& {Hrivnak}, B.~J. 1989, \apjl, 345, L51

\bibitem[{{Laor} \& {Draine}(1993)}]{Laor1993}
{Laor}, A. \& {Draine}, B.~T. 1993, \apj, 402, 441

\bibitem[{{Lebouteiller} {et~al.}(2011){Lebouteiller}, {Barry}, {Spoon},
  {Bernard-Salas}, {Sloan}, {Houck}, \& {Weedman}}]{Lebouteiller2011}
{Lebouteiller}, V., {Barry}, D.~J., {Spoon}, H.~W.~W., {et~al.} 2011, \apjs,
  196, 8

\bibitem[{{Matsuura} {et~al.}(2009){Matsuura}, {Barlow}, {Zijlstra},
  {Whitelock}, {Cioni}, {Groenewegen}, {Volk}, {Kemper}, {Kodama}, {Lagadec},
  {Meixner}, {Sloan}, \& {Srinivasan}}]{Matsuura2009}
{Matsuura}, M., {Barlow}, M.~J., {Zijlstra}, A.~A., {et~al.} 2009, \mnras, 396,
  918

\bibitem[{{Matsuura} {et~al.}(2014){Matsuura}, {Bernard-Salas}, {Lloyd Evans},
  {Volk}, {Hrivnak}, {Sloan}, {Chu}, {Gruendl}, {Kraemer}, {Peeters},
  {Szczerba}, {Wood}, {Zijlstra}, {Hony}, {Ita}, {Kamath}, {Lagadec}, {Parker},
  {Reid}, {Shimonishi}, {Van Winckel}, {Woods}, {Kemper}, {Meixner}, {Otsuka},
  {Sahai}, {Sargent}, {Hora}, \& {McDonald}}]{Matsuura2014}
{Matsuura}, M., {Bernard-Salas}, J., {Lloyd Evans}, T., {et~al.} 2014, \mnras,
  439, 1472

\bibitem[{{Matsuura} {et~al.}(2006){Matsuura}, {Wood}, {Sloan}, {Zijlstra},
  {van Loon}, {Groenewegen}, {Blommaert}, {Cioni}, {Feast}, {Habing}, {Hony},
  {Lagadec}, {Loup}, {Menzies}, {Waters}, \& {Whitelock}}]{Matsuura2006}
{Matsuura}, M., {Wood}, P.~R., {Sloan}, G.~C., {et~al.} 2006, \mnras, 371, 415

\bibitem[{{Meixner} {et~al.}(2006){Meixner}, {Gordon}, {Indebetouw}, {Hora},
  {Whitney}, {Blum}, {Reach}, {Bernard}, {Meade}, {Babler}, {Engelbracht},
  {For}, {Misselt}, {Vijh}, {Leitherer}, {Cohen}, {Churchwell}, {Boulanger},
  {Frogel}, {Fukui}, {Gallagher}, {Gorjian}, {Harris}, {Kelly}, {Kawamura},
  {Kim}, {Latter}, {Madden}, {Markwick-Kemper}, {Mizuno}, {Mizuno}, {Mould},
  {Nota}, {Oey}, {Olsen}, {Onishi}, {Paladini}, {Panagia}, {Perez-Gonzalez},
  {Shibai}, {Sato}, {Smith}, {Staveley-Smith}, {Tielens}, {Ueta}, {van Dyk},
  {Volk}, {Werner}, \& {Zaritsky}}]{Meixner2006}
{Meixner}, M., {Gordon}, K.~D., {Indebetouw}, R., {et~al.} 2006, \aj, 132, 2268

\bibitem[{{Meixner} {et~al.}(2013){Meixner}, {Panuzzo}, {Roman-Duval},
  {Engelbracht}, {Babler}, {Seale}, {Hony}, {Montiel}, {Sauvage}, {Gordon},
  {Misselt}, {Okumura}, {Chanial}, {Beck}, {Bernard}, {Bolatto}, {Bot},
  {Boyer}, {Carlson}, {Clayton}, {Chen}, {Cormier}, {Fukui}, {Galametz},
  {Galliano}, {Hora}, {Hughes}, {Indebetouw}, {Israel}, {Kawamura}, {Kemper},
  {Kim}, {Kwon}, {Lebouteiller}, {Li}, {Long}, {Madden}, {Matsuura}, {Muller},
  {Oliveira}, {Onishi}, {Otsuka}, {Paradis}, {Poglitsch}, {Reach},
  {Robitaille}, {Rubio}, {Sargent}, {Sewi{\l}o}, {Skibba}, {Smith},
  {Srinivasan}, {Tielens}, {van Loon}, \& {Whitney}}]{Meixner2013}
{Meixner}, M., {Panuzzo}, P., {Roman-Duval}, J., {et~al.} 2013, \aj, 146, 62

\bibitem[{{Min} {et~al.}(2013){Min}, {Jeffers}, {Canovas}, {Rodenhuis},
  {Keller}, \& {Waters}}]{Min2013}
{Min}, M., {Jeffers}, S.~V., {Canovas}, H., {et~al.} 2013, \aap, 554, A15

\bibitem[{{Ngeow} \& {Kanbur}(2008)}]{Ngeow2008}
{Ngeow}, C. \& {Kanbur}, S.~M. 2008, \apj, 679, 76

\bibitem[{{Nikutta} {et~al.}(2014){Nikutta}, {Hunt-Walker}, {Nenkova},
  {Ivezi{\'c}}, \& {Elitzur}}]{Nikutta2014}
{Nikutta}, R., {Hunt-Walker}, N., {Nenkova}, M., {Ivezi{\'c}}, {\v Z}., \&
  {Elitzur}, M. 2014, \mnras, 442, 3361

\bibitem[{{Ossenkopf} {et~al.}(1992){Ossenkopf}, {Henning}, \&
  {Mathis}}]{Ossenkopf1992}
{Ossenkopf}, V., {Henning}, T., \& {Mathis}, J.~S. 1992, \aap, 261, 567

\bibitem[{{Pegourie}(1988)}]{Pegourie1988}
{Pegourie}, B. 1988, \aap, 194, 335

\bibitem[{{Pietrzy{\'n}ski} {et~al.}(2013){Pietrzy{\'n}ski}, {Graczyk},
  {Gieren}, {Thompson}, {Pilecki}, {Udalski}, {Soszy{\'n}ski}, {Koz{\l}owski},
  {Konorski}, {Suchomska}, {Bono}, {Moroni}, {Villanova}, {Nardetto},
  {Bresolin}, {Kudritzki}, {Storm}, {Gallenne}, {Smolec}, {Minniti}, {Kubiak},
  {Szyma{\'n}ski}, {Poleski}, {Wyrzykowski}, {Ulaczyk}, {Pietrukowicz},
  {G{\'o}rski}, \& {Karczmarek}}]{Pietrzynski2013}
{Pietrzy{\'n}ski}, G., {Graczyk}, D., {Gieren}, W., {et~al.} 2013, \nat, 495,
  76

\bibitem[{{Reach} {et~al.}(2005){Reach}, {Megeath}, {Cohen}, {Hora}, {Carey},
  {Surace}, {Willner}, {Barmby}, {Wilson}, {Glaccum}, {Lowrance}, {Marengo}, \&
  {Fazio}}]{Reach2005}
{Reach}, W.~T., {Megeath}, S.~T., {Cohen}, M., {et~al.} 2005, \pasp, 117, 978

\bibitem[{{Riebel} {et~al.}(2012){Riebel}, {Srinivasan}, {Sargent}, \&
  {Meixner}}]{Riebel2012}
{Riebel}, D., {Srinivasan}, S., {Sargent}, B., \& {Meixner}, M. 2012, \apj,
  753, 71

\bibitem[{{Rieke} {et~al.}(2015){Rieke}, {Wright}, {B{\"o}ker}, {Bouwman},
  {Colina}, {Glasse}, {Gordon}, {Greene}, {G{\"u}del}, {Henning}, {Justtanont},
  {Lagage}, {Meixner}, {N{\o}rgaard-Nielsen}, {Ray}, {Ressler}, {van Dishoeck},
  \& {Waelkens}}]{Rieke2015}
{Rieke}, G.~H., {Wright}, G.~S., {B{\"o}ker}, T., {et~al.} 2015, \pasp, 127,
  584

\bibitem[{{Rieke} {et~al.}(2004){Rieke}, {Young}, {Engelbracht}, {Kelly},
  {Low}, {Haller}, {Beeman}, {Gordon}, {Stansberry}, {Misselt}, {Cadien},
  {Morrison}, {Rivlis}, {Latter}, {Noriega-Crespo}, {Padgett}, {Stapelfeldt},
  {Hines}, {Egami}, {Muzerolle}, {Alonso-Herrero}, {Blaylock}, {Dole}, {Hinz},
  {Le Floc'h}, {Papovich}, {P{\'e}rez-Gonz{\'a}lez}, {Smith}, {Su}, {Bennett},
  {Frayer}, {Henderson}, {Lu}, {Masci}, {Pesenson}, {Rebull}, {Rho}, {Keene},
  {Stolovy}, {Wachter}, {Wheaton}, {Werner}, \& {Richards}}]{Rieke2004}
{Rieke}, G.~H., {Young}, E.~T., {Engelbracht}, C.~W., {et~al.} 2004, \apjs,
  154, 25

\bibitem[{{Robitaille} {et~al.}(2006){Robitaille}, {Whitney}, {Indebetouw},
  {Wood}, \& {Denzmore}}]{Robitaille2006}
{Robitaille}, T.~P., {Whitney}, B.~A., {Indebetouw}, R., {Wood}, K., \&
  {Denzmore}, P. 2006, \apjs, 167, 256

\bibitem[{{Ruffle} {et~al.}(2015){Ruffle}, {Kemper}, {Jones}, {Sloan},
  {Kraemer}, {Woods}, {Boyer}, {Srinivasan}, {Antoniou}, {Lagadec}, {Matsuura},
  {McDonald}, {Oliveira}, {Sargent}, {Sewi{\l}o}, {Szczerba}, {van Loon},
  {Volk}, \& {Zijlstra}}]{Ruffle2015}
{Ruffle}, P.~M.~E., {Kemper}, F., {Jones}, O.~C., {et~al.} 2015, \mnras, 451,
  3504

\bibitem[{{Russell} \& {Dopita}(1992)}]{Russell1992}
{Russell}, S.~C. \& {Dopita}, M.~A. 1992, \apj, 384, 508

\bibitem[{{Sargent} {et~al.}(2011){Sargent}, {Srinivasan}, \&
  {Meixner}}]{Sargent2011}
{Sargent}, B.~A., {Srinivasan}, S., \& {Meixner}, M. 2011, \apj, 728, 93

\bibitem[{{Seale} {et~al.}(2009){Seale}, {Looney}, {Chu}, {Gruendl}, {Brandl},
  {Chen}, {Brandner}, \& {Blake}}]{Seale2009}
{Seale}, J.~P., {Looney}, L.~W., {Chu}, Y.-H., {et~al.} 2009, \apj, 699, 150

\bibitem[{{Sewi{\l}o} {et~al.}(2013){Sewi{\l}o}, {Carlson}, {Seale},
  {Indebetouw}, {Meixner}, {Whitney}, {Robitaille}, {Oliveira}, {Gordon},
  {Meade}, {Babler}, {Hora}, {Block}, {Misselt}, {van Loon}, {Chen},
  {Churchwell}, \& {Shiao}}]{Sewilo2013}
{Sewi{\l}o}, M., {Carlson}, L.~R., {Seale}, J.~P., {et~al.} 2013, \apj, 778, 15

\bibitem[{{Sheets} {et~al.}(2013){Sheets}, {Bolatto}, {van Loon}, {Sandstrom},
  {Simon}, {Oliveira}, \& {Barb{\'a}}}]{Sheets2013}
{Sheets}, H.~A., {Bolatto}, A.~D., {van Loon}, J.~T., {et~al.} 2013, \apj, 771,
  111

\bibitem[{{Sloan} {et~al.}(2015){Sloan}, {Herter}, {Charmandaris}, {Sheth},
  {Burgdorf}, \& {Houck}}]{Sloan2015a}
{Sloan}, G.~C., {Herter}, T.~L., {Charmandaris}, V., {et~al.} 2015, \aj, 149,
  11

\bibitem[{{Sloan} {et~al.}(2007){Sloan}, {Jura}, {Duley}, {Kraemer},
  {Bernard-Salas}, {Forrest}, {Sargent}, {Li}, {Barry}, {Bohac}, {Watson}, \&
  {Houck}}]{Sloan2007}
{Sloan}, G.~C., {Jura}, M., {Duley}, W.~W., {et~al.} 2007, \apj, 664, 1144

\bibitem[{{Sloan} {et~al.}(2016){Sloan}, {Kraemer}, {McDonald}, {Groenewegen},
  {Wood}, {Zijlstra}, {Lagadec}, {Boyer}, {Kemper}, {Matsuura}, {Sahai},
  {Sargent}, {Srinivasan}, {van Loon}, \& {Volk}}]{Sloan2016}
{Sloan}, G.~C., {Kraemer}, K.~E., {McDonald}, I., {et~al.} 2016, \apj, 826, 44

\bibitem[{{Sloan} {et~al.}(2008){Sloan}, {Kraemer}, {Wood}, {Zijlstra},
  {Bernard-Salas}, {Devost}, \& {Houck}}]{Sloan2008}
{Sloan}, G.~C., {Kraemer}, K.~E., {Wood}, P.~R., {et~al.} 2008, \apj, 686, 1056

\bibitem[{{Sloan} {et~al.}(2014){Sloan}, {Lagadec}, {Zijlstra}, {Kraemer},
  {Weis}, {Matsuura}, {Volk}, {Peeters}, {Duley}, {Cami}, {Bernard-Salas},
  {Kemper}, \& {Sahai}}]{Sloan2014}
{Sloan}, G.~C., {Lagadec}, E., {Zijlstra}, A.~A., {et~al.} 2014, \apj, 791, 28

\bibitem[{{Sloan} \& {Price}(1998)}]{SloanPrice1998}
{Sloan}, G.~C. \& {Price}, S.~D. 1998, \apjs, 119, 141

\bibitem[{{Smolders} {et~al.}(2010){Smolders}, {Acke}, {Verhoelst},
  {Blommaert}, {Decin}, {Hony}, {Sloan}, {Neyskens}, {van Eck}, {Zijlstra}, \&
  {van Winckel}}]{Smolders2010}
{Smolders}, K., {Acke}, B., {Verhoelst}, T., {et~al.} 2010, \aap, 514, L1

\bibitem[{{Speck} {et~al.}(2000){Speck}, {Barlow}, {Sylvester}, \&
  {Hofmeister}}]{Speck2000}
{Speck}, A.~K., {Barlow}, M.~J., {Sylvester}, R.~J., \& {Hofmeister}, A.~M.
  2000, \aaps, 146, 437

\bibitem[{{Srinivasan} {et~al.}(2016){Srinivasan}, {Boyer}, {Kemper},
  {Meixner}, {Sargent}, \& {Riebel}}]{Srinivasan2016}
{Srinivasan}, S., {Boyer}, M.~L., {Kemper}, F., {et~al.} 2016, \mnras, 457,
  2814

\bibitem[{{Srinivasan} {et~al.}(2011){Srinivasan}, {Sargent}, \&
  {Meixner}}]{Srinivasan2011}
{Srinivasan}, S., {Sargent}, B.~A., \& {Meixner}, M. 2011, \aap, 532, A54

\bibitem[{{Stanghellini} {et~al.}(2007){Stanghellini}, {Garc{\'{\i}}a-Lario},
  {Garc{\'{\i}}a-Hern{\'a}ndez}, {Perea-Calder{\'o}n}, {Davies}, {Manchado},
  {Villaver}, \& {Shaw}}]{Stanghellini2007}
{Stanghellini}, L., {Garc{\'{\i}}a-Lario}, P., {Garc{\'{\i}}a-Hern{\'a}ndez},
  D.~A., {et~al.} 2007, \apj, 671, 1669

\bibitem[{{Szewczyk} {et~al.}(2009){Szewczyk}, {Pietrzy{\'n}ski}, {Gieren},
  {Ciechanowska}, {Bresolin}, \& {Kudritzki}}]{Szewczyk2009}
{Szewczyk}, O., {Pietrzy{\'n}ski}, G., {Gieren}, W., {et~al.} 2009, \aj, 138,
  1661

\bibitem[{{Temim} {et~al.}(2015){Temim}, {Dwek}, {Tchernyshyov}, {Boyer},
  {Meixner}, {Gall}, \& {Roman-Duval}}]{Temim2015}
{Temim}, T., {Dwek}, E., {Tchernyshyov}, K., {et~al.} 2015, \apj, 799, 158

\bibitem[{{Ueta} \& {Meixner}(2003)}]{Ueta2003}
{Ueta}, T. \& {Meixner}, M. 2003, \apj, 586, 1338

\bibitem[{{Valiante} {et~al.}(2009){Valiante}, {Schneider}, {Bianchi}, \&
  {Andersen}}]{Valiante2009}
{Valiante}, R., {Schneider}, R., {Bianchi}, S., \& {Andersen}, A.~C. 2009,
  \mnras, 397, 1661

\bibitem[{{van Aarle} {et~al.}(2011){van Aarle}, {van Winckel}, {Lloyd Evans},
  {Ueta}, {Wood}, \& {Ginsburg}}]{vanAarle2011}
{van Aarle}, E., {van Winckel}, H., {Lloyd Evans}, T., {et~al.} 2011, \aap,
  530, A90

\bibitem[{{van der Marel} \& {Cioni}(2001)}]{vanderMarel2001}
{van der Marel}, R.~P. \& {Cioni}, M.-R.~L. 2001, \aj, 122, 1807

\bibitem[{{van Winckel}(2003)}]{VanWinckel2003}
{van Winckel}, H. 2003, \araa, 41, 391

\bibitem[{{Volk} {et~al.}(2011){Volk}, {Hrivnak}, {Matsuura}, {Bernard-Salas},
  {Szczerba}, {Sloan}, {Kraemer}, {van Loon}, {Kemper}, {Woods}, {Zijlstra},
  {Sahai}, {Meixner}, {Gordon}, {Gruendl}, {Tielens}, {Indebetouw}, \&
  {Marengo}}]{Volk2011}
{Volk}, K., {Hrivnak}, B.~J., {Matsuura}, M., {et~al.} 2011, \apj, 735, 127

\bibitem[{{Ward} {et~al.}(2017){Ward}, {Oliveira}, {van Loon}, \&
  {Sewi{\l}o}}]{Ward2017}
{Ward}, J.~L., {Oliveira}, J.~M., {van Loon}, J.~T., \& {Sewi{\l}o}, M. 2017,
  \mnras, 464, 1512

\bibitem[{{Whitney} {et~al.}(2008){Whitney}, {Sewilo}, {Indebetouw},
  {Robitaille}, {Meixner}, {Gordon}, {Meade}, {Babler}, {Harris}, {Hora},
  {Bracker}, {Povich}, {Churchwell}, {Engelbracht}, {For}, {Block}, {Misselt},
  {Vijh}, {Leitherer}, {Kawamura}, {Blum}, {Cohen}, {Fukui}, {Mizuno},
  {Mizuno}, {Srinivasan}, {Tielens}, {Volk}, {Bernard}, {Boulanger}, {Frogel},
  {Gallagher}, {Gorjian}, {Kelly}, {Latter}, {Madden}, {Kemper}, {Mould},
  {Nota}, {Oey}, {Olsen}, {Onishi}, {Paladini}, {Panagia}, {Perez-Gonzalez},
  {Reach}, {Shibai}, {Sato}, {Smith}, {Staveley-Smith}, {Ueta}, {Van Dyk},
  {Werner}, {Wolff}, \& {Zaritsky}}]{Whitney2008}
{Whitney}, B.~A., {Sewilo}, M., {Indebetouw}, R., {et~al.} 2008, \aj, 136, 18

\bibitem[{{Woods} {et~al.}(2011{\natexlab{a}}){Woods}, {Oliveira}, {Kemper},
  {van Loon}, {Sargent}, {Matsuura}, {Szczerba}, {Volk}, {Zijlstra}, {Sloan},
  {Lagadec}, {McDonald}, {Jones}, {Gorjian}, {Kraemer}, {Gielen}, {Meixner},
  {Blum}, {Sewi{\l}o}, {Riebel}, {Shiao}, {Chen}, {Boyer}, {Indebetouw},
  {Antoniou}, {Bernard}, {Cohen}, {Dijkstra}, {Galametz}, {Galliano}, {Gordon},
  {Harris}, {Hony}, {Hora}, {Kawamura}, {Lawton}, {Leisenring}, {Madden},
  {Marengo}, {McGuire}, {Mulia}, {O'Halloran}, {Olsen}, {Paladini}, {Paradis},
  {Reach}, {Rubin}, {Sandstrom}, {Soszy{\'n}ski}, {Speck}, {Srinivasan},
  {Tielens}, {van Aarle}, {van Dyk}, {van Winckel}, {Vijh}, {Whitney}, \&
  {Wilkins}}]{Woods2011}
{Woods}, P.~M., {Oliveira}, J.~M., {Kemper}, F., {et~al.} 2011{\natexlab{a}},
  \mnras, 411, 1597

\bibitem[{{Woods} {et~al.}(2011{\natexlab{b}}){Woods}, {Sloan}, {Gordon},
  {Shiao}, {Kemper}, {Indebetouw}, \& {for the SAGE-Spec team}}]{Woods2011b}
{Woods}, P.~M., {Sloan}, G.~C., {Gordon}, K.~D., {et~al.} 2011{\natexlab{b}},
  ArXiv e-prints

\bibitem[{{Wright} {et~al.}(2010){Wright}, {Eisenhardt}, {Mainzer}, {Ressler},
  {Cutri}, {Jarrett}, {Kirkpatrick}, {Padgett}, {McMillan}, {Skrutskie},
  {Stanford}, {Cohen}, {Walker}, {Mather}, {Leisawitz}, {Gautier}, {McLean},
  {Benford}, {Lonsdale}, {Blain}, {Mendez}, {Irace}, {Duval}, {Liu}, {Royer},
  {Heinrichsen}, {Howard}, {Shannon}, {Kendall}, {Walsh}, {Larsen}, {Cardon},
  {Schick}, {Schwalm}, {Abid}, {Fabinsky}, {Naes}, \& {Tsai}}]{Wright2010}
{Wright}, E.~L., {Eisenhardt}, P.~R.~M., {Mainzer}, A.~K., {et~al.} 2010, \aj,
  140, 1868

\bibitem[{{Wright} {et~al.}(2015){Wright}, {Wright}, {Goodson}, {Rieke},
  {Aitink-Kroes}, {Amiaux}, {Aricha-Yanguas}, {Azzollini}, {Banks},
  {Barrado-Navascues}, {Belenguer-Davila}, {Bloemmart}, {Bouchet}, {Brandl},
  {Colina}, {Detre}, {Diaz-Catala}, {Eccleston}, {Friedman},
  {Garc{\'{\i}}a-Mar{\'{\i}}n}, {G{\"u}del}, {Glasse}, {Glauser}, {Greene},
  {Groezinger}, {Grundy}, {Hastings}, {Henning}, {Hofferbert}, {Hunter},
  {Jessen}, {Justtanont}, {Karnik}, {Khorrami}, {Krause}, {Labiano}, {Lagage},
  {Langer}, {Lemke}, {Lim}, {Lorenzo-Alvarez}, {Mazy}, {McGowan}, {Meixner},
  {Morris}, {Morrison}, {M{\"u}ller}, {rgaard-Nielson}, {Olofsson},
  {O'Sullivan}, {Pel}, {Penanen}, {Petach}, {Pye}, {Ray}, {Renotte}, {Renouf},
  {Ressler}, {Samara-Ratna}, {Scheithauer}, {Schneider}, {Shaughnessy},
  {Stevenson}, {Sukhatme}, {Swinyard}, {Sykes}, {Thatcher}, {Tikkanen}, {van
  Dishoeck}, {Waelkens}, {Walker}, {Wells}, \& {Zhender}}]{Wright2015}
{Wright}, G.~S., {Wright}, D., {Goodson}, G.~B., {et~al.} 2015, \pasp, 127, 595

\bibitem[{{Zhang} {et~al.}(2009){Zhang}, {Jiang}, \& {Li}}]{Zhang2009}
{Zhang}, K., {Jiang}, B.~W., \& {Li}, A. 2009, \apj, 702, 680

\bibitem[{{Zubko} {et~al.}(1996){Zubko}, {Mennella}, {Colangeli}, \&
  {Bussoletti}}]{Zubko1996}
{Zubko}, V.~G., {Mennella}, V., {Colangeli}, L., \& {Bussoletti}, E. 1996,
  \mnras, 282, 1321

\end{thebibliography}


\end{document}